\documentclass[journal]{IEEEtran}
\pdfoutput=1
\usepackage{amsmath,bm}
\usepackage{amssymb}
\usepackage{ntheorem}
\usepackage{subfigure}
\usepackage{caption}
\usepackage{indentfirst}
\usepackage{ifpdf}
\usepackage{cite,color}
\usepackage{graphicx}
\usepackage{epstopdf}
\usepackage{array}
\usepackage{cite}
\usepackage{algorithm}
\usepackage{algorithmic}
\usepackage{multirow}
\usepackage{booktabs}
\usepackage{threeparttable}
\usepackage{diagbox}
\usepackage{float}
\usepackage{url}
\newenvironment{proof}{\noindent{\emph{Proof:}}}{\hfill$\square$}

\newcommand{\tabincell}[2]{\begin{tabular}{@{}#1@{}}#2\end{tabular}}

\begin{document}
\title{Vandermonde Factorization of Hankel Matrix for Complex Exponential Signal Recovery --- Application
in Fast NMR Spectroscopy}
\author{Jiaxi~Ying,
        Jian-Feng~Cai,
        Di~Guo,
        Gongguo~Tang,
        Zhong~Chen,
        Xiaobo~Qu*
\thanks{This work was supported by National Natural Science Foundation of China (61571380, 61811530021, 61871341, U1632274, 61672335 and 61601276), Natural Science Foundation of Fujian Province of China (2018J06018 and 2016J05205), Fundamental Research Funds for the Central Universities (20720150109), Science and Technology Program of Xiamen (3502Z20183053) and Hong Kong Research Grant Council (16300616). Currently this paper has been accepted on IEEE Transactions on Signal Processing. (*Corresponding author: Xiaobo Qu)}
\thanks{Jiaxi Ying, Zhong Chen and Xiaobo Qu are with the
Department of Electronic Science, Fujian Provincial Key Laboratory of Plasma and Magnetic Resonance, Xiamen University, Xiamen, China (e-mail:
 quxiaobo@xmu.edu.cn).}
\thanks{Jian-Feng Cai is with Department of Mathematics, Hong Kong University
of Science and Technology, Hong Kong SAR, China.}
\thanks{Di Guo is with the School of Computer and Information Engineering, Xiamen University of Technology, Xiamen, China.}
\thanks{Gongguo Tang is with Department of Electrical Engineering \& Computer Science, Colorado School of Mines, Golden, CO, United States.}}

\maketitle

\begin{abstract}
Many signals are modeled as a superposition of exponential functions in spectroscopy of chemistry, biology and medical imaging. This paper studies the problem of recovering exponential signals from a random subset of samples. We exploit the Vandermonde structure of the Hankel matrix formed by the exponential signal and formulate signal recovery as Hankel matrix completion with Vandermonde factorization (HVaF). A numerical algorithm is developed to solve the proposed model and its sequence convergence is analyzed theoretically. Experiments on synthetic data demonstrate that HVaF succeeds over a wider regime than the state-of-the-art nuclear-norm-minimization-based Hankel matrix completion method, while has a less restriction on frequency separation than the state-of-the-art atomic norm minimization and fast iterative hard thresholding methods. The effectiveness of HVaF is further validated on biological magnetic resonance spectroscopy data.
\end{abstract}


\IEEEpeerreviewmaketitle
\section{Introduction}
\IEEEPARstart{S}{ignals} in many practical applications can be exactly or approximately modeled as a superposition of a few complex exponential functions. Examples include analog-to-digital conversion in electronic systems \cite{Tropp2010}, antenna signals in tele-communication \cite{Nion2010,qian2017puma}, images in fluorescence microscopy \cite{schermelleh2010}, echo signals in medical imaging \cite{liang1989high} and time domain signals in nuclear magnetic resonance (NMR) spectroscopy \cite{Qu-Accelerate-2015, ying2017, nguyen2013denoising,lu2018low,Guo2017fast,ye2017localised}. The signal of interest in these applications is a superposition of a few complex sinusoids with or without damping factors:
\begin{equation}\label{eq:signal}
y(t) = \sum\limits_{r = 1}^R {{c_r}{e^{(2\pi i{f_r} - {\tau _r})t}}} ,
\end{equation}
where  ${f_r} \in [0,1)$ is the normalized frequency, ${c_r} \in {\mathbb{C}}$ is the associated complex amplitude, $R$ is the number of exponentials and  ${\tau _r} \in {\mathbb{R}_ + }$ is the damping factor. Particularly, when there is no damping factor, i.e., ${\tau _r} = 0$ for $r = 1, \ldots ,R$, $y(t)$ is a spectrally sparse signal which arises from, e.g., analog-to-digital conversion \cite{Tropp2010}; otherwise, $ y(t) $ is a sum of damped complex sinusoids that characters the time domain signal, e.g. the acquired signal in biological NMR spectroscopy \cite{Qu-Accelerate-2015}.

Throughout the paper, it is assumed that the frequencies ${f_r}$, $r=1,\ldots,R$, are distinct and normalized with respect to Nyquist frequency, thus measurements are sampled at integer values. Let $\bm{y} \in {\mathbb {C}^{2N - 1}}$ be the underlying uniformly-sampled true signal $\bm{y} = [y(1) \ \ldots \ y(2N-1)]^T$.

In some circumstances, the measurements of the signal ${\bm{y}}$ are incomplete due to high experimental cost, hardware limitation, or other inevitable reasons. For example, to accelerate data acquisition, non-uniform sampling is popular in NMR spectroscopy \cite{Qu-Accelerate-2015}. In this paper, we aim to recover ${\bm{y}}$ from partial measurements $\left\{ {{y_j},j \in \Omega } \right\}$, where $\Omega  \subset \{ 1, \ldots ,2N - 1\} $ with $\left| \Omega  \right| = M$ $(M < 2N - 1)$ is the set of indices of the observed entries. Since the number of degrees of freedom to determine ${\bm{y}}$ is much smaller than the ambient dimension $2N-1$ if $R \ll 2N - 1$, it is possible to recover ${\bm{y}}$ from a small number of measurements by exploiting its inherent structure.

One line of work tries to exploit sparsity of $\bm{y}$ in the frequency domain. Spectrally sparse signals, i.e., sums of complex sinusoids without damping factors, enjoy sparse representations in the discrete Fourier transform domain if frequencies are aligned well with the discrete frequencies. In that case, the signal can be recovered from a minimal number of samples by using conventional compressed sensing \cite{Candes2006}. However, true frequencies often take values in the continuous domain, and the resultant basis mismatch between the true frequencies and the discretized grid \cite{Chi-Mismatch-2011} leads to loss of sparsity and hence degrades the performance of compressed sensing. Therefore, total variation or atomic norm \cite{Chandrasekaran2012} minimization methods were proposed to address this problem by exploiting the sparsity of $\bm{y}$ with continuous-valued frequencies \cite{Tang2013, Candes-SuperResolution-2014}. Exact recovery with high probability is established for random sampling of very few measurements, provided that the frequencies ${f_r}$, $1, \ldots ,R$, enjoy good separations \cite{Tang2013}. Several subsequent papers on this topic include \cite{chi2015compressive, yang2015gridless, xu2014precise,fang2014super,Costa2017}. Yet, currently it is still unknown how to extend atomic norm minimization to recover damped exponential signals.

Recently, the low rank structure of the Hankel matrix formed by $\bm{y}$ is exploited to recover the signal \cite{fazel2013hankel, Markovsky2013, Andersson-FreqEstim-2014, Chen-Chi2014, Qu-Accelerate-2015, Cai-RobustLR-2016, usevich2016hankel,Cai-Wirtinger-2015,markovsky2014recent, Cai-FIHT-2016,cai2017spectral}, and the reconstruction problem can be formulated under the framework of  Low Rank Hankel Matrix Completion (LRHMC),
\begin{equation}\label{eq:LRHMC}
\min_{\bm{x}} \mathrm{rank}(H(\bm{x})),
\ s.t.\
\mathcal{P}_\Omega(\bm{x}) = \mathcal{P}_\Omega(\bm{y}),
\end{equation}
where $H(\bm{x})$ denotes a Hankel matrix arranged from $\bm x$; $\mathcal{P}_\Omega$ is the orthogonal projector onto the indices in $\Omega $ so that the $j$-th element of $ \mathcal{P}_\Omega (\bm{y})$ is equal to ${y_j}$ if $j \in \Omega $ and zero otherwise.

There are several existing algorithms in the literature to solve LRHMC problem \eqref{eq:LRHMC}. In particular, nuclear norm minimization is utilized as a convex relaxation for LRHMC and theoretical recovery guarantees are established \cite{Chen-Chi2014,Cai-RobustLR-2016}. This approach was independently developed to recover time domain signals in biological NMR spectroscopy \cite{Qu-Accelerate-2015}, showing great potentials to highly accelerate data acquisition with non-uniform sampling. To design more efficient algorithms, some non-convex algorithms for low-rank Hankel matrix completion were proposed in \cite{Cai-FIHT-2016,cai2017spectral}. However, some critical issues remain to be solved, e.g., the recovery is sensitive to frequency separation \cite{Cai-FIHT-2016, cai2017spectral} and the number of measurements are expected to be further reduced in fast sampling \cite{Qu-Accelerate-2015, Chen-Chi2014, Cai-RobustLR-2016}. In addition, low rank Hankel matrix formulation has been extended to reconstruct higher-dimensional NMR spectroscopy with block Hankel matrix \cite{lu2018low, Guo2017fast} and tensors \cite{ying2017}. Applications of low rank Hankel matrix can also be found in magnetic resonance imaging \cite{haldar2014,ongie2016off,jin2016general,shin2014calibrationless}.

Note that the signal of interest in this paper is a special signal function discussed in the theory of finite rate of innovation (FRI) \cite{vetterli2002sampling}. The class of FRI signals includes a stream of (differentiated) Diracs, non-uniform splines and piecewise smooth polynomials. The conditions for perfect reconstruction from low pass filtered observation were established under ideal low-pass filters and gaussian kernels \cite{vetterli2002sampling}, or kernels satisfying Strange-Fix conditions \cite{dragotti2007sampling}. The reconstruction scheme estimates an annihilating filter that annihilates the Fourier series coefficients of an FRI signal at consecutive low frequencies. More recently, a unified view for sampling and reconstruction in the frequency domain was proposed \cite{haro2018sampling} to deal with arbitrary sampling kernels. Some robust methods were also developed to handle the noisy situations \cite{haro2018sampling, blu2008sparse,condat2015cadzow}. However, these methods are not originally designed for missing data recovery and thus some modifications are necessary in this situation.

In this work, we exploit the Vandermonde structure of Hankel matrix to recover complex exponential signals. A new numerical algorithm is developed to implement this approach and its convergence properties are further analysed. The extensive simulations demonstrate that, compared with state-of-the-art methods, the new approach requires fewer measurements to recover signals. Moreover, signal parameters can be accurately estimated through HVaF reconstruction even with a small frequency separation. The advantages of the proposed approach are further validated on biological NMR spectroscopy data. A preliminary account of this study was presented in a recent conference paper \cite{Qu2017}.

It is worth noting that this paper focuses on signal reconstruction rather than parameter estimation which actually is a harmonic retrieval problem. To better clarify the contributions of this work, the main differences between this work and the work \cite{ying2017} are summarized here. First, the signal processing problem to be solved and their applications are different. This work is to improve the fundamental 1-dimensional complex exponential signal reconstruction and is highly motivated by fast sampling in 2-dimensional NMR spectroscopy \cite{Qu-Accelerate-2015} while \cite{ying2017} was used in $N$-dimensional ($N\geq 3$) spectroscopy. In addition, the new method may be extended to other applications such as magnetic resonance imaging and hyperspectral imaging, where signals in one dimension can be modelled as sum of exponentials. Second, reconstruction models are different. This work first explores Vandermonde structure of Hankel matrix in 1-dimensional exponential signal reconstruction, while the work \cite{ying2017} exploits the natural tensor structures of the $N$-dimensional exponential signal. Third, this work makes new and important contributions to signal processing. The new method achieves higher empirical phase transition and enjoys a less restriction on frequency separation, compared with the state-of-the-art spectrally sparse signal reconstruction methods \cite{Tang2013, Chen-Chi2014, Cai-FIHT-2016, Qu-Accelerate-2015}.

The rest of this paper is organized as follows. Section \ref{sec:II} introduces notations. In Section \ref{sec:III}, we propose the recovery method and a numerical algorithm to implement it. Section \ref{sec:IV} presents the numerical results on both synthetic and biological NMR spectroscopy data. Section \ref{sec:V} extends discussions on the proposed method and Section \ref{sec:VI} finally concludes this work and discusses future directions.

\section{Notations}\label{sec:II}
We first introduce the notation used throughout this paper. We denote vectors by bold lowercase letters and matrices by bold uppercase letters. The individual entries of vectors and matrices are denoted by normal font. More explicitly, the $j$-th entry of a vector ${\bm{x}}$ is denoted by ${x_j}$; the $(i,j)$-th entry of a matrix ${\bm{X}}$ is denoted by ${X_{ij}}$. The $i$-th row and $j$-th column of a matrix $\bm{X}$ are denoted by $\bm{X}_{(i,:)}$ and $\bm{X}_{(:,j)}$, respectively. For any matrix $\bm{X}$, ${\left\| \bm{X} \right\|_*}$ and ${\left\| \bm{X} \right\|_F}$ denote nuclear norm and Frobenius norm, respectively. The transpose of vectors and matrices are denoted by $\bm{x}^T$ and $\bm{X}^T$, while their conjugate transpose is denoted by $\bm{x}^H$ and $\bm{X}^H$. The Hadamard product (also known as entrywise product) of two matrices ${\bm{A}}$ and ${\bm{B}}$ is $[\bm{A} \odot \bm{B}]_{ij} = A_{ij} B_{ij}$.

Operators are denoted by calligraphic letters. Let $\mathcal{R}$ be a Hankel operator which maps a vector $\bm{x} \in \mathbb{C}^n$ to a Hankel matrix $\mathcal{R} \bm{x} \in {\mathbb{C}^{{n_1} \times {n_2}}}$ with ${n_1} + {n_2} = n + 1$ as follows
\begin{equation}\label{eq:OpHank}
\mathcal{R}:{\bm{x}} \in {\mathbb{C}^{{n_1} + {n_2} - 1}} \mapsto {\kern 1pt} \mathcal{R} \bm{x} \in \mathbb{C}^{{n_1} \times {n_2}},[\mathcal{R}{\bm{x}}]_{ij} = {x_{i + j-1}},
\end{equation}
and the adjoint operator $\mathcal{R}^*$ of $\mathcal{R}$ is given by
\begin{gather}\label{eq:AdOpHank}
\mathcal{R}^*:{\kern 1pt} \bm{X} \in {\mathbb{C}^{{n_1} \times {n_2}}} \mapsto \mathcal{R}^* {\bm{X}} \in {\mathbb{C}^{{n_1} + {n_2} - 1}}, \\
[\mathcal{R}^* \bm{X}]_k=\sum\limits_{i + j-1= k}{X_{ij}}, \notag
\end{gather}
for any $i \in \{ 1, \ldots ,n_1 \}$ and $j \in \{ 1, \ldots ,n_2 \}$. In particular, we denote the Hankel operator by $\mathcal{H}$ instead of $\mathcal{R}$ when ${n_1} = {n_2}$, i.e., constructing a square matrix.

We denote $\mathcal{G} = \mathcal{R}^*\mathcal{R}$. Obviously, $\mathcal{G}$ is a diagonal operator satisfying $\mathcal{G} \bm{x} = \bm{Wx}$, $\bm{x} \in {\mathbb{C}^{{n_1} + {n_2} - 1}}$, where ${\bm{W}}$ is a diagonal matrix and its element on the main diagonal is the number of times that an entry of vector ${\bm{x}}$ is presented in the Hankel matrix. The Moore-Penrose pseudoinverse of $\mathcal{R}$ is denoted by $\mathcal{R}^\dag  = \mathcal{G}^{ - 1}\mathcal{R}^*$ which satisfies ${\mathcal{R}^\dag } \mathcal{R} = \mathcal{I}$, where $\mathcal{I}$ is the identity operator.

We also define another linear operator $\mathcal{Q}_r$ which aims to extract the $r$-th column from $\bm{X}$. For a matrix $\bm{X} \in {\mathbb{C}^{{s_1} \times {s_2}}}$, specifically, we define $\mathcal{Q}_r$ by
\begin{equation}\label{eq:OpQ}
\mathcal{Q}_r:{\kern 1pt} \bm{X} \in {\mathbb{C}^{{s_1} \times {s_2}}}{\kern 3pt}  \mapsto \mathcal{Q}_r \bm{X} = \bm{X}_{(:,r)} \in {\mathbb{C}^{{s_1} \times 1}},
\end{equation}
for any $ r \in \{ 1, \ldots , s_2 \}$.

Then the adjoint $\mathcal{Q}_r^*$ of $\mathcal{Q}_r$ is given by
\begin{gather}\label{eq:AdOpQ}
\mathcal{Q}_r^*:\bm{x} \in {\mathbb{C}^{{s_1} \times 1}} \mapsto \mathcal{Q}_r^*{\bm{x}} \in {\mathbb{C}^{{s_1} \times {s_2}}},\\
\left[ \mathcal{Q}_r^*\bm{x} \right]_{(:,k)} = \left\{ {\begin{array}{*{20}{c}}
\bm{x}\\
\bm{0}
\end{array}\begin{array}{*{20}{c}}
{}\\
{}
\end{array}\begin{array}{*{20}{c}}
{k = r,}\\
{k \ne r.}
\end{array}} \right. \quad \forall r \in \{ 1, \ldots ,s_2\} \notag.
\end{gather}
where ${\left[  \cdot  \right]_{(:,k)}}$ denotes the $k$-th column of a matrix, $k = 1, \ldots , s_2$. Thus we have
\begin{equation}
\left[ \mathcal{Q}_r^*\mathcal{Q}_r{\bm{X}} \right]_{(:,k)} = \left\{ {\begin{array}{*{20}{c}}
{{\bm{X}}_{(:,r)}}\\
\bm{0}
\end{array}{\kern 10pt}  \begin{array}{*{20}{c}}
{k = r,}\\
{k \ne r.}
\end{array}} \right.
\end{equation}

\section{The proposed method}\label{sec:III}
In this section, we first propose a new approach for exponential signal recovery. The proposed approach formulates the reconstruction problem as Hankel matrix completion with Vandermonde factorization. Then we develop a numerical algorithm to implement the new approach. Finally we analyse the convergence properties of the algorithm.

\subsection{Hankel matrix completion with Vandermonde factorization}\label{sec:VF}
Define $z_r= e^{(2\pi i{f_r} - {\tau _r})}$ for  $r = 1, \ldots ,R$. It is observed that the Hankel matrix $\mathcal{H} \bm{y} \in {\mathbb{C}^{N \times N}}$ formed by the signal of interest $\bm{y} \in {\mathbb{C}^{2N - 1}}$ in \eqref{eq:signal} admits Vandermonde factorization
\begin{equation}\label{eq:VandF}
\begin{array}{l}
\mathcal{H} \bm{y} = \left[ {\begin{array}{*{20}{c}}
1& \cdots &1\\
{{z_1}}& \cdots &{{z_R}}\\
 \vdots & \vdots & \vdots \\
{z_1^{N - 1}}& \cdots &{z_R^{N - 1}}
\end{array}} \right]\left[ {\begin{array}{*{20}{c}}
{{c_1}}&{}&{}\\
{}& \ddots &{}\\
{}&{}&{{c_R}}
\end{array}} \right]\\
{\kern 60pt}\left[ {\begin{array}{*{20}{c}}
1&{{z_1}}& \cdots &{z_1^{N - 1}}\\
 \vdots & \vdots & \cdots & \vdots \\
1&{{z_R}}& \cdots &{z_R^{N - 1}}
\end{array}} \right]= \bm{E\Sigma}\bm{E}^T,
\end{array}
\end{equation}
where $\bm{E}$ is a Vandermone matrix and $\bm{\Sigma}$ is a diagonal matrix. We rewrite \eqref{eq:VandF} as a form of the product of two factor matrices
\begin{equation}\label{eq:TVandF}
\mathcal{H}\bm{y} = \bm{E}_L\bm{E}_R^T,
\end{equation}
where $\bm{E}_L = \bm{E}\bm{\Sigma }_L$ and $\bm{E}_R = \bm{E}\bm{\Sigma}_R$ with $\bm{\Sigma}_L\bm{\Sigma }_R^T = \bm{\Sigma}$, and $\bm{\Sigma}_L$ and $\bm{\Sigma }_R \in {\mathbb{C}^{R \times R}}$ are both diagonal matrices. Clearly, \eqref{eq:TVandF} can be easily converted to \eqref{eq:VandF} by normalizing $\bm{E}_L$ and $\bm{E}_R$. Therefore, for ease of presentation in this paper, we also call \eqref{eq:TVandF} as Vandermonde factorization and $\bm{E}_L$ and $\bm{E}_R$ are with Vandermonde structure in this paper. In addition, structured matrix factorization was also explored in applications such as blind separations \cite{chan2008convex} and power spectra separations\cite{fu2016power}.

In this paper, we reconstruct complex exponential signal by imposing Vandermonde factorization on Hankel matrix:
\begin{gather}
\mathrm{Find} \ \bm{x},\bm{U},\bm{V} \label{eq:OriVand}\\
\quad\mbox{subject to}\quad
\bm{U} \ \mathrm{and} \ \bm{V} \ \mbox{are of Vandermonde structure},\notag\\
\mathcal{H}\bm{x} = \bm{U}\bm{V}^T,{\kern 3pt} \mathcal{P}_\Omega(\bm{x}) = \mathcal{P}_\Omega(\bm{y}).\notag
\end{gather}
Specifically, \eqref{eq:OriVand} aims to find $\bm{x}$, which is consistent with $\bm{y}$ in $\Omega $ and the Hankel matrix formed by which can be factorized into two matrices with Vandermonde structure. From \eqref{eq:VandF}, it is observed that Vandermonde factorization is a special form of low rank matrix factorization; the product of factor matrices from Vandermonde factorization compose a low rank matrix, meanwhile they comply with Vandermonde structure. Therefore, \eqref{eq:OriVand} imposes more constraints in recovery than LRHMC in \eqref{eq:LRHMC} and thus has a potential to achieve a better reconstruction.

Unfortunately, it is hard to directly impose Vandermonde structure on $\bm{U}$ and $\bm{V}$, since the set of all Vandermonde matrices is non-convex and highly nonlinear, which may cause the optimization problem to be computationally intractable. In this paper, Vandermonde structure is expected by seeking each column of the matrix to be an exponential function.

Note that $\mathrm{rank}(\mathcal{R}\bm{a})=1$ if $\bm{a}$ is a column of a matrix with Vandermonde structure. Let $\bm{A}=[\bm{a}_1,\ldots,\bm{a}_R]\in\mathbb{C}^{N\times R}$ be a matrix without zero columns. It is obvious that $R\leq \sum_{r=1}^{R}\mathrm{rank}(\mathcal{R}\bm{a}_r)$. More importantly, $R=\sum_{r=1}^{R}\mathrm{rank}(\mathcal{R}\bm{a}_r)$ if and only if $\bm{A}$ is Vandermonde. Therefore, minimizing $\sum_{r=1}^{R}\mathrm{rank}(\mathcal{R}\bm{a}_r)$ will favor $\bm{A}$ to be of Vandermonde structure. Based on this observation, we propose
\begin{gather}
\min_{\bm{U},\bm{V},\bm{x}} \sum\limits_{r = 1}^{\hat R} {(\mathrm{rank}(\mathcal{R}\bm{U}_{(:,r)})+\mathrm{rank}( \mathcal{R}\bm{V}_{(:,r)} )}),\label{eq:Model1}\\
s.t. \quad \mathcal{H}\bm{x} = \bm{U}\bm{V}^T,\mathcal{P}_\Omega(\bm{x}) = \mathcal{P}_\Omega(\bm{y}).\notag
\end{gather}
The objective function is to encourage $\bm{U}$ and $\bm{V}$ are with Vandermonde structure. However, due to the rank function involved, it is hard to solve the above minimization. Here we relax rank function by nuclear norm, and solve
\begin{gather}
\min_{\bm{U},\bm{V},\bm{x}} \sum\limits_{r = 1}^{\hat R} {(\left\| \mathcal{R}\bm{U}_{(:,r)} \right\|_*+\left\| \mathcal{R}\bm{V}_{(:,r)} \right\|_*}),\label{eq:Model2}\\
s.t. \quad \mathcal{H}\bm{x} = \bm{U}\bm{V}^T,\mathcal{P}_\Omega(\bm{x}) = \mathcal{P}_\Omega(\bm{y}).\notag
\end{gather}

Here both $\mathcal{R}$ and $\mathcal{H}$ are defined to map a vector to a Hankel matrix. By the operator $\mathcal{H}$, the 1-D exponential signal recovery problem is reformulated as Hankel matrix completion which has been explored in \cite{Chen-Chi2014, Qu-Accelerate-2015, usevich2016hankel, markovsky2014recent, Cai-FIHT-2016}. To further impose the Vandermonde structure which is first explored in our paper, we map each column of $\bm{U}$ and $\bm{V}$ to Hankel matrix by using operator $\mathcal{R}$ and then minimize the nuclear norm of each Hankel matrix. Therefore, using two different notations helps present our contributions more clearly.

We adopt the type of $\mathcal{H}\bm{x} =\bm{U}\bm{V}^T$ rather than symmetrical decomposition form, which is shown in \eqref{eq:VandF}, with using $\bm U$ only due to the fact that the Hankel matrix may not be exactly square when the dimension of $\bm{x}$ is even. Therefore, our proposed approach is not limited to odd dimensional signals. In addition, using the former type can find the closed form solutions of $\bm U$ and $\bm V$ directly but not for the latter.

A numerical algorithm is developed in the next subsection and the effectiveness is verified by experiments on synthetic data and realistic biological magnetic resonance spectroscopy data in Section \ref{sec:IV}.

Note that the exact number of exponentials $R$ is usually unknown in practice especially in the case of non-uniform sampling, hence our algorithm needs to preset the number of exponentials. It will be shown in Section \ref{sec:IV} that the proposed algorithm can succeed in recovering the signal with high probability even when the preset number of exponentials $\hat R$ is greatly larger than $R$.

\subsection{Numerical algorithm}\label{sec:model}
In this subsection, we develop a numerical algorithm to implement the proposed method. Recall that ${{\cal Q}_r}$ defined in \eqref{eq:OpQ}. We rewrite \eqref{eq:Model2} as its equivalence
\begin{gather}
\min_{\bm{U},\bm{V},\bm{x}} \sum\limits_{r = 1}^{\hat R} {(\left\| \mathcal{R}{\mathcal{Q}_r}\bm{U} \right\|_*+\left\| \mathcal{R}{\mathcal{Q}_r}\bm{V}\right\|_*)},\label{eq:Model4}\\
s.t.\ \mathcal{H}\bm{x} = \bm{U}\bm{V}^T,\mathcal{P}_\Omega (\bm{x}) = \mathcal{P}_\Omega(\bm{y}). \notag
\end{gather}

To solve \eqref{eq:Model4}, we develop an algorithm based on half quadratic methods with continuation \cite{Geman1995, Nikolova2005} for its advantage in handling multi-variable optimization \cite{Qu2014}. We introduce the term $\left\| \mathcal{H}\bm{x} - \bm{U}\bm{V}^T \right\|_F^2$ to keep $\mathcal{H}\bm{x}$ close enough to $\bm{U}\bm{V}^T$ instead of the exact constraint $\mathcal{H}\bm{x} = \bm{U}\bm{V}^T$. Now, we have the following optimization:
\begin{gather}
\min_{\bm{U},\bm{V},\bm{x}} \sum\limits_{r = 1}^{\hat{R}} {({\left\| \mathcal{R}{{\cal Q}_r}\bm{U} \right\|}_*
+ \left\| {{\cal R}{{\cal Q}_r}\bm{V}} \right\|_*)}+\frac{\beta }{2}\left\| {\mathcal{H}\bm{x} - \bm{U}\bm{V}^T} \right\|_F^2,  \notag \\
s.t.\ \mathcal{P}_\Omega(\bm{x}) = \mathcal{P}_\Omega(\bm{y}).\label{eq:Model3}
\end{gather}
When $\beta \to \infty $, the solution to \eqref{eq:Model3} is approaching the one to \eqref{eq:Model4}. The challenge in solving \eqref{eq:Model3} is that the nuclear norm terms are non-smooth and non-separable simultaneously. To decouple the non-smoothness and the non-separability, we introduce some auxiliary variables into \eqref{eq:Model3} \cite{fang2016FDLCP}, and then reformulate the problem \eqref{eq:Model3} as the following equivalent form:
\begin{gather}
\min_{\substack{ \bm{U},\bm{V},\bm{x}, \bm{B}_{r}, \bm{C}_{r}\\r=1, \cdots,\hat{R} }} \sum\limits_{r = 1}^{\hat{R}} {({\left\| \bm{B}_r \right\|}_* + {\left\| \bm{C}_r \right\|}_*)} +\frac{\beta }{2}\left\| {\mathcal{H}\bm{x} - \bm{U}{\bm{V}}^T} \right\|_F^2 , \notag \\
s.t.\ \bm{B}_r = \mathcal{R}\mathcal{Q}_r \bm{U},\ \bm{C}_r= \mathcal{R}\mathcal{Q}_r\bm{V},\ r \in \{ 1, \ldots ,{\hat R}\},\label{eq:AuxiMode}\\
\mathcal{P}_\Omega (\bm{x}) = {\mathcal{P}_\Omega }(\bm{y}). \notag
\end{gather}

In \eqref{eq:AuxiMode}, the first two terms are non-smooth but separable, and the other terms are smooth, which makes it easier to develop numerical algorithms. With a fixed $\beta $, we use Alternating Direction Method of Multipliers (ADMM) to solve \eqref{eq:AuxiMode}.

In the following, we first present an overview of the proposed algorithm, and then describe how to update each variable of the algorithm in detail. Let $\mathcal{B} = (\bm{B}_1, \ldots ,\bm{B}_{\hat R})$, $\mathcal{C} = (\bm{C}_1, \ldots ,\bm{C}_{\hat R})$, $\mathcal{D} = (\bm{D}_1, \ldots ,\bm{D}_{\hat R})$ and $\mathcal{M} = (\bm{M}_1, \ldots ,\bm{M}_{\hat R})$. The augmented Lagrangian function of \eqref{eq:AuxiMode} is
\begin{equation}
\begin{array}{l} \label{eq:LargModel}
\mathcal{L}_\mu (\bm{U},\bm{V},\bm{x},\mathcal{B},\mathcal{D},\mathcal{C},\mathcal{M}) = \frac{\beta }{2}\left\| {{\cal H}\bm{x} - \bm{U}\bm{V}^T} \right\|_F^2\\
+\sum\limits_{r = 1}^{\hat{R}} {(\left\| \bm{B}_r \right\|_* + \left\langle {\bm{D}_r,\mathcal{R}\mathcal{Q}_r\bm{U} - \bm{B}_r} \right\rangle  + \frac{\mu }{2}\left\| {\mathcal{R}{\mathcal{Q}_r}\bm{U} - \bm{B}_r} \right\|_F^2)} \\
+\sum\limits_{r = 1}^{\hat{R}} {(\left\| \bm{C}_r \right\|_* + \left\langle {\bm{M}_r,\mathcal{R}\mathcal{Q}_r\bm{V} - \bm{C}_r} \right\rangle  + \frac{\mu }{2}\left\| \mathcal{R}\mathcal{Q}_r\bm{V} - \bm{C}_r \right\|_F^2)} .
\end{array}
\end{equation}
where $\bm{D}_r$ and $\bm{M}_r$ are the Lagrange multipliers, $r=1,\cdots, \hat R$.

We present an ADMM iterative scheme to successively minimize $\mathcal{L}_\mu(\bm{U},\bm{V},\bm{x},\mathcal{B},\mathcal{C},\mathcal{D},\mathcal{M})$ as follows:

\begin{flalign}
&\min_{\bm{U}} {\mathcal{L}_{{\mu}^k}}(\bm{U},\bm{V}^k,\bm{x}^k,\mathcal{B}^k,\mathcal{C}^k,\mathcal{D}^k,\mathcal{M}^k) \label{eq:SolveU}\\
&\min_{\bm{V}} {\mathcal{L}_{{\mu}^k}}(\bm{U}^{k + 1},\bm{V},\bm{x}^k,\mathcal{B}^k,\mathcal{C}^k,\mathcal{D}^k,\mathcal{M}^k) \label{eq:SolveV}\\
&\min_{\bm{x}} {\mathcal{L}_{{\mu}^k}}(\bm{U}^{k + 1},\bm{V}^{k + 1},\bm{x},\mathcal{B}^k,\mathcal{C}^k,\mathcal{D}^k,\mathcal{M}^k) \label{eq:Solvex}\\
&{\kern 50pt} s.t. \ \mathcal{P}_\Omega(\bm{x}) = \mathcal{P}_\Omega(\bm{y}). \nonumber\\
&\min_{\mathcal{B}} {\mathcal{L}_{\mu ^k}}({\bm{U}}^{k + 1},\bm{V}^{k + 1},\bm{x}^{k + 1},\mathcal{B},\mathcal{C}^k,\mathcal{D}^k,\mathcal{M}^k)\label{eq:SolveB}\\
&\min_{\mathcal{C}} {\mathcal{L}_{\mu ^k}}(\bm{U}^{k + 1},\bm{V}^{k + 1},\bm{x}^{k + 1},\mathcal{B}^{k + 1},\mathcal{C},\mathcal{D}^k,\mathcal{M}^k) \label{eq:SolveC}\\
&\bm{D}_r^{k + 1} = \bm{D}_r^k + \mu ^k(\mathcal{R}{\mathcal{Q}_r}\bm{U}^{k + 1} - \bm{B}_r^{k + 1}) \label{eq:SolveD}\\
&{\kern 50pt}\forall r \in \{ 1, \ldots ,{\hat R}\}.\nonumber\\
&\bm{M}_r^{k + 1} = \bm{M}_r^k + \mu ^k(\mathcal{R}{\mathcal{Q}_r}\bm{V}^{k + 1} - \bm{C}_r^{k + 1}) \label{eq:SolveM}\\
&{\kern 50pt}\forall r \in \{ 1, \ldots ,{\hat R}\}.\nonumber
\end{flalign}

\vspace{2ex}
\noindent 1) Update $\bm{U}$ and $\bm{V}$

To update the variable $\bm{U}$, the optimization problem \eqref{eq:SolveU} is written as follows:
\begin{equation}\label{eq:UpdatU}
\begin{array}{l}
\min\limits_{\bm{U}} \frac{\beta }{2}\left\| {\mathcal{H}\bm{x}^k - \bm{U}\left( \bm{V}^k \right)^T} \right\|_F^2 \\
{\kern 20pt}+\frac{\mu^k}{2}\sum\limits_{r = 1}^{\hat R} {\left\| \mathcal{R}\mathcal{Q}_r\bm{U} - \bm{B}_r^k + (\mu ^k)^{ - 1}\bm{D}_r^k \right\|_F^2}.
\end{array}
\end{equation}
Since \eqref{eq:UpdatU} is a least square problem, its solution is obtained by solving a system of linear equations as follows
\begin{equation}\label{eq:SolU}
\begin{array}{l}
\mu ^k\sum\limits_{r = 1}^{\hat R} {\mathcal{Q}_r^*\mathcal{R}^*\mathcal{R}\mathcal{Q}_r\bm{U}}  + \beta \bm{U}(\bm{V}^k)^T\mathrm{conj}(\bm{V}^k)\\
= \sum\limits_{r = 1}^{\hat R} {\mu ^k\mathcal{Q}_r^*{\mathcal{R}^*}\left( \bm{B}_r^k - \bm{D}_r^k/\mu ^k \right)}  + \beta \left( \mathcal{H}\bm{x}^k \right)\mathrm{conj}(\bm{V}^k),
\end{array}
\end{equation}
where $\mathrm{conj}(\cdot)$ denotes the conjugate of a matrix.

Similar to the case of $\bm{U}$, we can update $\bm{V}$ by solving the following equation
\begin{equation}\label{eq:UpdatV}
\begin{array}{l}
\min\limits_{\bm{V}} \frac{\beta }{2}\left\| \mathcal{H}\bm{x}^k - \bm{U}^{k + 1}\bm{V}^T \right\|_F^2\\
{\kern 20pt}+\frac{\mu ^k}{2}\sum\limits_{r = 1}^{\hat R} {\left\| \mathcal{R}\mathcal{Q}_r\bm{V} - \bm{C}_r^k + (\mu ^k)^{ - 1}\bm{M}_r^k \right\|_F^2},
\end{array}
\end{equation}
and its solution can be obtained by
\begin{equation}\label{eq:SolV}
\begin{array}{l}
\mu ^k\sum\limits_{r = 1}^{\hat R} {\mathcal{Q}_r^*\mathcal{R}^*\mathcal{R}\mathcal{Q}_r\bm{V}}  + \beta\bm{V}(\bm{U}^{k + 1})^T\mathrm{conj}(\bm{U}^{k + 1}) \\
= \sum\limits_{r = 1}^{\hat R} {\mu ^k\mathcal{Q}_r^*\mathcal{R}^*\left( \bm{C}_r^k - \bm{M}_r^k/\mu ^k \right)}  + \beta \left( \mathcal{H}\bm{x}^k \right)^T\mathrm{conj}(\bm{U}^{k + 1}).
\end{array}
\end{equation}
The specific closed-form solution of $\bm{U}$ and $\bm{V}$ are given in appendix.

\vspace{2ex}
\noindent 2) Update $\bm{x}$

We update the variable $\bm{x}$ by solving
\begin{equation}\label{eq:Updatx}
\min \limits_{\bm{x}} \left\| \mathcal{H}\bm{x} - \bm{U}^{k + 1}{\left( \bm{V}^{k + 1} \right)}^T \right\|_F^2, \ s.t. \ {\mathcal{P}_\Omega }(\bm{x}) = {\mathcal{P}_\Omega }(\bm{y}).
\end{equation}

By introducing Lagrangian multiplier $\bm{d}$ for the constraint $\mathcal{P}_\Omega (\bm{x}) = \mathcal{P}_\Omega (\bm{y})$, we write the Lagrangian function of \eqref{eq:Updatx} as follows:
\begin{equation}\label{eq:Larx}
F(\bm{x},\bm{d}) = \left\| \mathcal{H}\bm{x} - \bm{U}^{k + 1}{\left( \bm{V}^{k + 1} \right)}^T \right\|_F^2 + \left\langle \bm{d},\mathcal{P}_\Omega (\bm{x}) - \mathcal{P}_\Omega (\bm{y}) \right\rangle.
\end{equation}

By setting ${\nabla_{(\bm{x},\bm{d})}}F = \bm{0}$, we have the Karush\text{-}Kuhn\text{-}Tucker (KKT) conditions
\[\left( \mathcal{H}^*\mathcal{H}\bm{x} - \mathcal{H}^*\left( \bm{U}^{k + 1}\left( \bm{V}^{k + 1} \right)^T \right) \right) - \mathcal{P}_\Omega(\bm{d}) = \bm{0}.\]
\[\mathcal{P}_\Omega(\bm{x}) - \mathcal{P}_\Omega(\bm{y}) = \bm{0}.\]

By deriving the KKT conditions simply, one has
\begin{equation}\label{eq:Solux}
\bm{x}^{k + 1} = \mathcal{P}_\Omega(\bm{y}) + \mathcal{P}_{\Omega ^\mathrm{C}}\left( \bm{W}^{ - 1}\left( \mathcal{H}^*\left( \bm{U}^{k + 1}\left( \bm{V}^{k + 1} \right)^T \right) \right) \right).
\end{equation}
where $\Omega^\mathrm{C}$ is the complement of $\Omega $, i.e., the set of indices of the unobserved entries, and $\bm{W}$ is a constant matrix as introduced in Section \ref{sec:II}.

\vspace{2ex}
\noindent 3) Update $\bm{B}_r$ and $\bm{C}_r$

To update $\bm{B}_r$, we solve the following sub-problem,
\begin{equation}\label{eq:UpdatB}
\begin{aligned}
&\min \limits_{\bm{B}_r} \| \bm{B}_r \|_* + \left\langle \bm{D}_r^k, \mathcal{R}\mathcal{Q}_r\bm{U}^{k + 1} - \bm{B}_r \right\rangle  \\
&{\kern 70pt}+ \frac{{\mu ^k}}{2}\left\| \mathcal{R}\mathcal{Q}_r\bm{U}^{k + 1} - \bm{B}_r \right\|_F^2.
\end{aligned}
\end{equation}
Following  \cite{Cai2010}, the closed-form solution of \eqref{eq:UpdatB} is
\begin{equation}\label{eq:SoluB}
\bm{B}_r^{k+1} = S_{1/{\mu ^k}}\left( \mathcal{R}\mathcal{Q}_r\bm{U}^{k + 1} + 1/{\mu ^k}\bm{D}_r^k \right),
\end{equation}
where $S$ is the soft singular value thresholding operator \cite{Cai2010} with threshold $1/{\mu ^k}$. Similar to update $\bm{B}_r$, we can update $\bm{C}_r$ by
\begin{equation}\label{eq:UpdatC}
\begin{aligned}
&\min \limits_{\bm{C}_r} \left\| \bm{C}_r \right\|_* + \left\langle \bm{M}_r^k, \mathcal{R}\mathcal{Q}_r\bm{V}^{k + 1} - \bm{C}_r \right\rangle \\
&{\kern 70pt}+ \frac{\mu^k}{2}\left\| \mathcal{R}\mathcal{Q}_r\bm{V}^{k + 1} - \bm{C}_r \right\|_F^2,
\end{aligned}
\end{equation}
and its solution is
\begin{equation}\label{eq:SoluC}
\bm{C}_r^{k+1} = S_{1/{\mu ^k}}\left( \mathcal{R}\mathcal{Q}_r\bm{V}^{k + 1} + 1/{\mu ^k}\bm{M}_r^k \right)
\end{equation}

For a fixed nonzero $\beta $, the solution to \eqref{eq:Model3} only yields an approximation to the solution to \eqref{eq:Model1}. To obtain a better solution, we apply a continuation scheme in which we gradually improve $\beta $ to $+\infty $. This algorithm can also be accelerated by adaptively changing $\mu$. An efficient strategy \cite{Shang2016,Lin2009} is to increase $\mu^k$ iteratively by $\mu ^{k + 1} = \rho \mu ^k$, where $\rho  \in (1,\ 1.1]$. The overall algorithm is summarized in Algorithm \ref{alg:1}.

In real applications, the measurements are usually contaminated by Gaussian noise. To recover the signal from noisy measurements, we propose the following optimization
\begin{equation}\label{eq:NoisyModel}
\begin{array}{l}
\min \limits_{\bm{U},\bm{V},\bm{x}} \sum\limits_{r = 1}^{\hat R} (\left\| \mathcal{R}\mathcal{Q}_r\bm{U} \right\|_* + \left\| \mathcal{R}\mathcal{Q}_r\bm{V} \right\|_*)  + \frac{\beta }{2}\left\| \mathcal{H}\bm{x} - \bm{U}\bm{V}^T \right\|_F^2\\
{\kern 80pt} + \frac{\lambda }{2}\left\| \mathcal{P}_\Omega (\bm{x}) - \mathcal{P}_\Omega (\bm{y}) \right\|_2^2.
\end{array}
\end{equation}

All the variables except $\bm{x}$ can be updated as above. We update the variable $\bm{x}$ by
\begin{equation*}
\min \limits_{\bm{x}} \frac{\beta }{2}\left\| \mathcal{H}\bm{x} - \bm{U}^{k + 1}\left( \bm{V}^{k + 1} \right)^T \right\|_F^2 + \ \frac{\lambda }{2}\left\| \mathcal{P}_\Omega (\bm{x}) - \mathcal{P}_\Omega (\bm{y}) \right\|_2^2,
\end{equation*}
and thus
\begin{equation}\label{eq:SoluNoisy}
\begin{array}{l}
\bm{x}^{k + 1} = \left( \beta \bm{W} + \lambda \mathcal{P}_\Omega ^*\mathcal{P}_\Omega \right)^{ - 1}\\
{\kern 20pt}\left( \beta \left( \mathcal{H}^*\left( \bm{U}^{k + 1}\left( \bm{V}^{k + 1} \right)^T \right) \right) + \lambda \mathcal{P}_\Omega^*\mathcal{P}_\Omega(\bm{y}) \right).
\end{array}
\end{equation}

The variable $\bm{x}$ is updated by \eqref{eq:SoluNoisy} in recovering the realistic biological NMR spectroscopy data as shown in Section \ref{sec:NMR}.

\begin{algorithm}[ht]
\caption{Algorithm for Hankel matrix completion with Vandermonde factorization (HVaF)\label{alg:1}}
\begin{algorithmic}[1]
\STATE Initialization: $\bm{U}$, $\bm{V}$, $\bm{x} = \mathcal{P}_\Omega (\bm{y})$ , $\beta  = {2^5}$ , $\beta_{\max } = 2^{30}$, $\mu ^0=10^{-2}$, and $\rho=1.05$.
\WHILE {$\beta  \le {\beta_{\max }}$}
\STATE Update $\bm{U}$, $\bm{V}$ by solving \eqref{eq:SolU} and \eqref{eq:SolV};
\STATE Update $\bm{x}$ by solving \eqref{eq:Solux};
\STATE For $r$=1 to $\hat R$, update $\bm{B}_r$ and $\bm{C}_r$ by solving \eqref{eq:SoluB} and \eqref{eq:SoluC};
\STATE For $r$=1 to $\hat R$, update $\bm{D}_r$ and $\bm{M}_r$ by solving \eqref{eq:SolveD} and \eqref{eq:SolveM};
\STATE Update $\mu $ by $\mu ^{k + 1} = \rho \mu ^k$;
\STATE If $\left\| \Delta \bm{x} \right\| = {\left\| \bm{x}_\mathrm{last} - \bm{x} \right\|} \mathord{\left/
{\vphantom {\left\| \bm{x}_\mathrm{last} - \bm{x} \right\|_2 \left\| \bm{x}_\mathrm{last} \right\|_2}} \right.
\kern-\nulldelimiterspace} \left\| \bm{x}_{\mathrm{last}} \right\| > 10^{-7} $, $\bm{x}_\mathrm{last} \leftarrow \bm{x}$, go to step 3; otherwise, go to step 9;
\STATE $\beta  \leftarrow 2\beta $;
\ENDWHILE
\end{algorithmic}
\end{algorithm}

\subsection{Convergence Analysis}
In this section, we analyse the convergence of Algorithm \ref{alg:1}. In Theorem \ref{thm1}, we show that the sequences $\{ \bm{U}^k\} $, $\{ \bm{V}^k\} $, $\{ \bm{x}^k\} $, $\{ \mathcal{B}^k\} $ and $\{ \mathcal{C}^k\} $ generated by Algorithm \ref{alg:1} with fixed $\beta$ converge.

\newtheorem{thm1}{Theorem}
\begin{thm1}\label{thm1}
The sequences $\{ \bm{U}^k\} $, $\{ \bm{V}^k\} $, $\{ \bm{x}^k\} $, $\{ \mathcal{B}^k\} $ and $\{ \mathcal{C}^k \} $ generated by  Algorithm \ref{alg:1} with fixed $\beta$ are all convergent.
\end{thm1}

Before proving the convergence properties of the algorithm, we first prove the boundedness of multipliers and some variables generated by Algorithm \ref{alg:1}.

\newtheorem{lem1}{Lemma}
\begin{lem1}\label{lem1}
The sequences $\{ \bm{D}_r^k\} $ and $\{ \bm{M}_r^k\} $, $r = 1, \ldots ,{\hat R}$, are bounded, where $\bm{D}_r^{k + 1} = \bm{D}_r^k + \mu ^k(\mathcal{R}\mathcal{Q}_r\bm{U}^{k + 1} - \bm{B}_r^{k + 1})$ and $\bm{M}_r^{k + 1} = \bm{M}_r^k + \mu ^k(\mathcal{R}\mathcal{Q}_r\bm{V}^{k + 1} - \bm{C}_r^{k + 1})$.
\end{lem1}

\begin{proof}
The optimality condition of \eqref{eq:UpdatB} gives
$$
\bm{0} \in \partial \left\| \bm{B}_r^{k + 1} \right\|_* - \bm{D}_r^k - \mu ^k(\mathcal{R}\mathcal{Q}_r\bm{U}^{k + 1} - \bm{B}_r^{k + 1}),
$$
which combined with \eqref{eq:SolveD} implies
\begin{equation*}
\bm{D}_r^{k + 1} \in \partial \left\| \bm{B}_r^{k + 1} \right\|_*.
\end{equation*}
According to \cite{Cai2010}, each element of the subgradient of the nuclear norm is bounded by 1 in spectral norm. Therefore ${\left\| \bm{D}_r^{k + 1} \right\|_2} \le 1$ and hence the sequence $\{ \bm{D}_r^k\} $ is bounded for all $r \in \{ 1, \ldots ,{\hat R}\} $. The boundedness of $\{ \bm{M}_r^k\} $ can be proved similarly.
\end{proof}

\newtheorem{lem2}[lem1]{Lemma}
\begin{lem2}\label{lem2}
The sequences $\{ \bm{U}^k\} $, $\{ \bm{V}^k\} $, $\{ \bm{x}^k\} $, $\{ \mathcal{B}^k\} $ and $\{ \mathcal{C}^k\} $ produced by Algorithm \ref{alg:1} with fixed $\beta$ are all bounded.
\end{lem2}

\begin{proof}
The augmented Lagrangian function satisfies
\begin{equation*}
\begin{aligned}
&{\kern 10pt}\mathcal{L}_{\mu^k}(\bm{U}^{k + 1},\bm{V}^{k + 1},\bm{x}^{k + 1},\mathcal{B}^{k + 1},\mathcal{C}^{k + 1},\mathcal{D}^{k + 1},\mathcal{M}^k)\\
&\le \mathcal{L}_{\mu ^k}(\bm{U}^k,\bm{V}^k,\bm{x}^k,\mathcal{B}^k,\mathcal{C}^k,\mathcal{D}^k,\mathcal{M}^k)\\
&= \mathcal{L}_{\mu ^{k - 1}}(\bm{U}^k,\bm{V}^k,\bm{x}^k,\mathcal{B}^k,\mathcal{C}^k,\mathcal{D}^k,\mathcal{M}^{k - 1})\\
&{\kern 10pt}+ \sum\limits_{r = 1}^{\hat R} {\left\langle \bm{M}_r^k - \bm{M}_r^{k - 1},\mathcal{R}\mathcal{Q}_r\bm{V}^k - \bm{C}_r^k \right\rangle }\\
&{\kern 10pt}+ \frac{\mu^k - \mu^{k - 1}}{2}\left\| \mathcal{R}\mathcal{Q}_r \bm{V}^k - \bm{C}_r^k \right\|_F^2\\
&= \mathcal{L}_{\mu ^{k - 1}}(\bm{U}^k,\bm{V}^k,\bm{x}^k,\mathcal{B}^k,\mathcal{C}^k,\mathcal{D}^k,\mathcal{M}^{k - 1}) \\
&{\kern 10pt}+ \frac{\mu ^k + \mu^{k - 1}}{2(\mu^{k - 1})^2}\sum\limits_{r = 1}^{\hat R} {\left\| \bm{M}_r^k - \bm{M}_r^{k - 1} \right\|_F^2} .
\end{aligned}
\end{equation*}
Summing it over $k$ gives
\begin{equation}\label{eq:proof1}
\begin{aligned}
&\mathcal{L}_{\mu^k}(\bm{U}^{k + 1},\bm{V}^{k + 1},\bm{x}^{k + 1},\mathcal{B}^{k + 1},\mathcal{C}^{k + 1},\mathcal{D}^{k + 1},\mathcal{M}^k)\\
&\le \mathcal{L}_{\mu ^0}(\bm{U}^1,\bm{V}^1,\bm{x}^1,\mathcal{B}^1,\mathcal{C}^1,\mathcal{D}^1,\mathcal{M}^0)\\
&\ + \sum_{j=1}^{k}\left(\frac{\mu ^j + \mu^{j - 1}}{2(\mu^{j - 1})^2}\sum\limits_{r = 1}^{\hat R} {\left\| \bm{M}_r^j - \bm{M}_r^{j- 1} \right\|_F^2}\right)\\
&\le \mathcal{L}_{\mu ^0}(\bm{U}^1,\bm{V}^1,\bm{x}^1,\mathcal{B}^1,\mathcal{C}^1,\mathcal{D}^1,\mathcal{M}^0)\\
&\ + \left(\sum_{j=1}^{k}\frac{\mu ^j + \mu^{j - 1}}{2(\mu^{j - 1})^2}\right)\left(\max_{j=1}^{k}\sum\limits_{r = 1}^{\hat R} {\left\| \bm{M}_r^j - \bm{M}_r^{j- 1} \right\|_F^2}\right).
\end{aligned}
\end{equation}
Since $\{ \bm{M}_r^k\}$ is bounded, $\max_{j=1}^{\infty}\sum_{r = 1}^{\hat R} {\left\| \bm{M}_r^j - \bm{M}_r^{j- 1} \right\|_F^2}$ is bounded. Furthermore, because of $\mu^{k+1}=\rho\mu^k$ and $\rho  \in (1,\ 1.1]$,
\begin{equation*}
\sum\limits_{j = 1}^\infty  {\frac{\mu ^j + \mu ^{j - 1}}{2(\mu^{j - 1})^2} = \frac{\rho (\rho  + 1)}{2\mu ^0(\rho  - 1)}}  < \infty.
\end{equation*}
Hence, $\{ \mathcal{L}_{\mu ^{k - 1}}(\bm{U}^k,\bm{V}^k,\bm{x}^k,\mathcal{B}^k,\mathcal{C}^k,\mathcal{D}^k,\mathcal{M}^{k - 1})\} $ is bounded. Similarly,
$\{ \mathcal{L}_{\mu ^{k - 1}}(\bm{U}^k,\bm{V}^k,\bm{x}^k,\mathcal{B}^k,\mathcal{C}^k,\mathcal{D}^{k - 1},\mathcal{M}^{k - 1})\} $ is also bounded.
We further have that
\begin{equation*}
\begin{aligned}
&\frac{\beta }{2}\left\| \mathcal{H}\bm{x}^k - \bm{U}^k\left( \bm{V}^k \right)^T \right\|_F^2 + \sum\limits_{r = 1}^{\hat R} {(\left\| \bm{B}_r^k \right\|_* + \left\| \bm{C}_r^k \right\|_*)} \\
&= \mathcal{L}_{\mu ^{k - 1}}(\bm{U}^k,\bm{V}^k,\bm{x}^k,\mathcal{B}^k,\mathcal{C}^k,\mathcal{D}^{k-1},\mathcal{M}^{k-1})\\
&-\frac{1}{\mu ^{k - 1}}\sum\limits_{r = 1}^{\hat R} {\big( \langle \bm{D}_r^{k - 1}, \bm{D}_r^k - \bm{D}_r^{k - 1} \rangle  + \frac{1}{2}\left\| \bm{D}_r^k - \bm{D}_r^{k - 1} \right\|_F^2 \big)} \\
&-\frac{1}{\mu ^{k - 1}}\sum\limits_{r = 1}^{\hat R} {\big( {\langle \bm{M}_r^{k - 1}, \bm{M}_r^k - \bm{M}_r^{k - 1} \rangle  + \frac{1}{2}\left\| \bm{M}_r^k - \bm{M}_r^{k - 1} \right\|_F^2} \big)}.
\end{aligned}
\end{equation*}
Since all terms on the right hand side are bounded, the left hand side is bounded. Thus, $\{ {{\cal B}^k}\} $ and $\{ {{\cal C}^k}\} $ are bounded.

Finally, we have $\mathcal{R}\mathcal{Q}_r\bm{U}^{k + 1} = \bm{B}_r^{k + 1} + (\bm{D}_r^{k + 1} - \bm{D}_r^k)/\mu ^k$ and $\mathcal{R}\mathcal{Q}_r\bm{V}^{k + 1} = \bm{C}_r^{k + 1} + (\bm{M}_r^{k + 1} - \bm{M}_r^k)/\mu^k$, $r = 1, \ldots ,{\hat R}$. Also, $\mathcal{R}\mathcal{Q}_r$ is injective. Therefore, the sequences $\{ \bm{U}^k\} $ and $\{ \bm{V}^k\} $ are also bounded. According to \eqref{eq:Solux}, $\{ \bm{x}^k\} $ is bounded, too.
\end{proof}

Now we are ready to prove Theorem \ref{thm1}.

\noindent \emph{Proof of Theorem \ref{thm1}:}\\
The updating formula \eqref{eq:SolveD}
and the optimality condition of \eqref{eq:UpdatU} with respect to ${{\bf{U}}^{k+1}}$ imply
\begin{equation*}
\begin{aligned}
&\sum\limits_{r = 1}^{\hat R} \left( \mu ^k\mathcal{Q}_r^*\mathcal{R}^*\mathcal{R}\mathcal{Q}_r\bm{U}^{k + 1} - \mu ^k\mathcal{Q}_r^*\mathcal{R}^*\bm{B}_r^k + \mathcal{Q}_r^*\mathcal{R}^*\bm{D}_r^k \right )\\
&\ + \beta \bm{U}^{k + 1}\left( \bm{V}^k \right)^T\mathrm{conj}(\bm{V}^k) - \beta (\mathcal{R}\bm{x}^k)\mathrm{conj}(\bm{V}^k)\\
&=\mu ^k\sum\limits_{r = 1}^{\hat R} \mathcal{Q}_r^*\mathcal{R}^*({\cal R}\mathcal{Q}_r(\bm{U}^{k + 1} - \bm{U}^k)+\frac{1}{\mu ^k}\bm{D}_r^k- \frac{1}{\mu ^{k - 1}}\bm{D}_r^{k - 1}) \\
&\ + \mu ^k\sum\limits_{r = 1}^{\hat R} \mathcal{Q}_r^*\mathcal{R}^*(\mathcal{R}\mathcal{Q}_r\bm{U}^k - \bm{B}_r^k + \frac{1}{\mu ^{k - 1}}\bm{D}_r^{k - 1} ) \\
&\ + \beta \bm{U}^{k + 1}\left( \bm{V}^k \right)^T\mathrm{conj}(\bm{V}^k) - \beta (\mathcal{R}\bm{x}^k)\mathrm{conj}(\bm{V}^k)= \bm{0}.
\end{aligned}
\end{equation*}
Since $\sum\limits_{r = 1}^{\hat R} \mathcal{Q}_r^*\mathcal{R}^*\mathcal{R}\mathcal{Q}_r(\bm{U}^{k + 1} - \bm{U}^k)$ can be denoted by $\bm{T} \odot (\bm{U}^{k + 1} - \bm{U}^k)$, where $\bm{T}$ is a constant matrix (the specific form of $\bm{T}$ can be seen in the appendix) and $ \odot $ denotes Hadamard product, we have
\begin{equation*}
\begin{aligned}
&{\kern 10pt}\bm{T} \odot (\bm{U}^{k + 1} - \bm{U}^k) \\
&= \frac{\beta}{\mu ^k}\left((\mathcal{R}\bm{x}^k)\mathrm{conj}(\bm{V}^k) - \bm{U}^{k + 1}\left( \bm{V}^k \right)^T\mathrm{conj}(\bm{V}^k)\right) \\
&-\frac{1}{\mu ^k}\sum\limits_{r = 1}^{\hat R} \mathcal{Q}_r^*\mathcal{R}^*((\rho  + 1)\bm{D}_r^k - \rho \bm{D}_r^{k - 1}) := \frac{\bm{G}^k}{\mu ^k}.
\end{aligned}
\end{equation*}
By Lemma \ref{lem2}, the sequence $\{\bm{G}_k\}$ is bounded, and denote $\delta$ its upper bound. We have $\|\bm{G}_k\|_F\leq\delta$ for all $k$. Then, for any $m$ and any $n\geq m$,
\begin{equation*}
\begin{aligned}
&{\left\| \bm{T} \odot (\bm{U}^n - \bm{U}^m) \right\|_F} \le {\left\| \bm{T} \odot (\bm{U}^n - \bm{U}^{n - 1}) \right\|_F} \\
&+ {\left\| \bm{T} \odot (\bm{U}^{n - 1} - \bm{U}^{n - 2}) \right\|_F} +  \ldots  + \left\| \bm{T} \odot (\bm{U}^{m + 1} - \bm{U}^m) \right\|_F\\
&= \frac{\left\| \bm{G}^{n - 1} \right\|_F}{\mu ^{n - 1}} + \frac{\left\| \bm{G}^{n - 2} \right\|_F}{\mu ^{n - 2}} +  \ldots  + \frac{\left\| \bm{G}^m \right\|_F}{\mu ^m} \\
&\le \frac{\delta }{\mu ^m}\left( {\frac{1}{\rho^{n - m - 1}} + \frac{1}{\rho^{n - m - 2}} +  \ldots  + 1} \right) \leq\frac{\delta \rho}{\mu^m(\rho-1)},
\end{aligned}
\end{equation*}
Since $\frac{\delta \rho }{\mu ^m( \rho-1 )} \to 0$, $\{ \bm{T} \odot\bm{U}^k\} $ is a Cauchy sequence. Therefore, by the fact that $ \bm{T}$ has no zero entry, $\{\bm{U}^k\} $ is a Cauchy sequence and hence convergent. Similarly, $\{ \bm{V}^k\} $ is also convergent. With this, since by \eqref{eq:Solux} $\bm{x}^k$ is a continuous function of $\bm{U}^k$ and $\bm{V}^k$, the sequence $\{\bm{x}^k\}$ is convergent.

It remains to show the boundedness of $\{ \mathcal{B}^k\}$ and $\{ \mathcal{C}^k\}$. By \eqref{eq:SolveD}, we have $\bm{B}_r^{k + 1}=\mathcal{R}\mathcal{Q}_r\bm{U}^{k + 1}-(\bm{D}_r^{k + 1} - \bm{D}_r^k)/\mu ^k$. Furthermore, $\{\bm{U}^k\}$ is convergent, $\{\bm{D}_r^k\}$ is bounded, and $\lim_{k\to\infty}\mu^k=\infty$. Therefore, $\{\bm{B}_r^k\}$ is convergent for each $r \in \{ 1, \ldots ,{\hat R} \}$ and then $\{ \mathcal{B}^k\} $ is convergent. Similarly, $\{ \mathcal{C}^k\} $ is also convergent.
\hfill$\square$

\subsection{Computational complexity}\label{sec:complxty}
The main running time of the algorithm is dominated by performing singular value decomposition (SVD) for the singular value thresholding operator. Consider to recover a signal ${\bm x} \in \mathbb C^{2N-1}$ with the number of estimated exponentials $\hat R$. The SVD is performed on the Hankel matrix with the size of $0.5N \times 0.5N$, and thus the total time complexity of SVD in each iteration is $O({\hat R} N^3)$.

\section{Numerical experiments}\label{sec:IV}
In this section, we evaluate the performance of the proposed HVaF on synthetic data and realistic biological NMR spectroscopy data. For synthetic data, the HVaF is compared with three state-of-the-art algorithms, ANM \cite{Tang2013}, EMaC \cite{Chen-Chi2014}, FIHT \cite{Cai-FIHT-2016}. ANM and EMaC are implemented using CVX \cite{Michael2014}. For the realistic NMR data, the HVaF is compared with the state-of-the-art LRHM method \cite{Qu-Accelerate-2015}. All the compared methods are conducted using publicly available codes with default parameters. Here the comparisons with the method in \cite{cai2017spectral} is omitted because  the method in \cite{cai2017spectral} together with EMaC and FIHT still belongs to the framework of LRHMC where the complex exponential signal recovery problem is formulated as low rank Hankel matrix completion.

\subsection{Synthetic data}\label{sec:Phase}

\begin{figure*}[ht]
\centering
\includegraphics[width=6.5in]{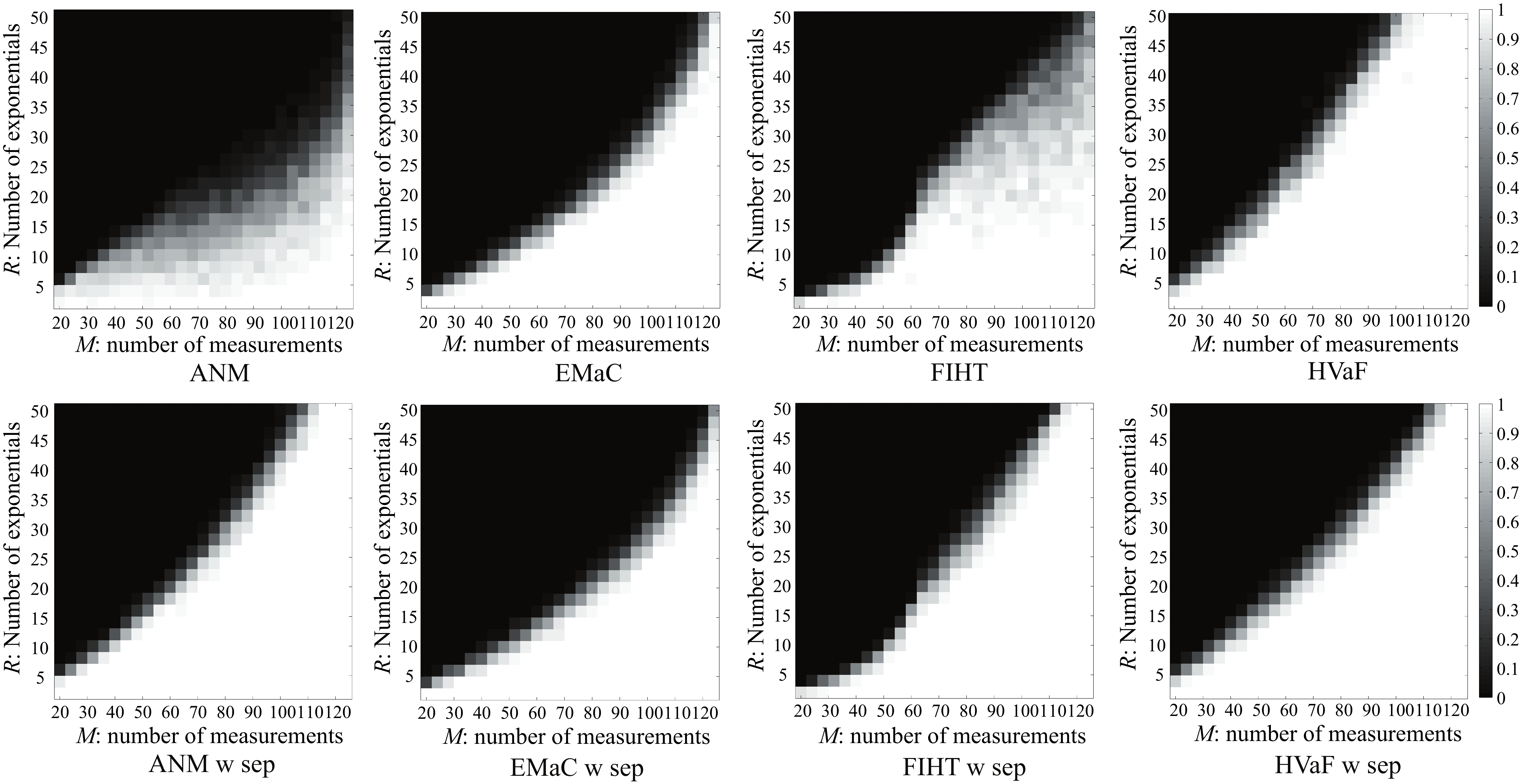}
\caption{Phase transition of successful reconstruction on undamped signals. Top: no restriction on frequencies of test signals; Bottom: wrap-around distances between frequencies are at least 1.5/$(2N-1)$. The length of the test signal $2N-1$ is 127.}
\label{fig1}
\end{figure*}

\begin{figure*}[ht]
\centering
\includegraphics[width=5.7in]{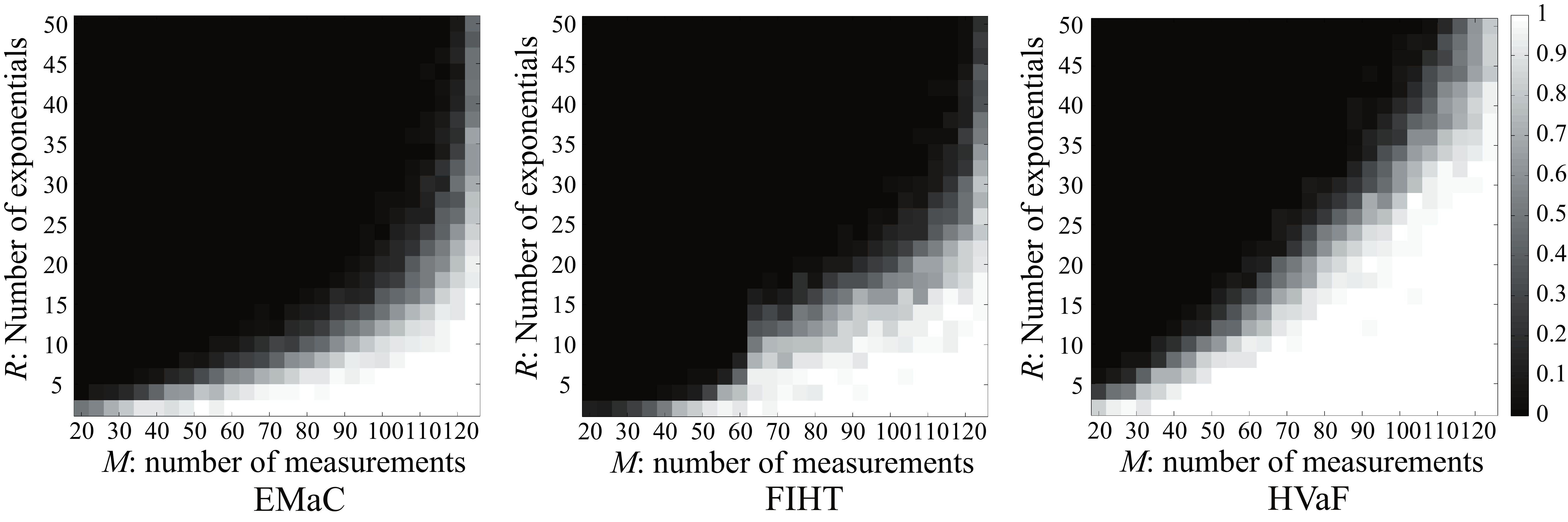}
\caption{Phase transition of successful reconstruction on damped signals. The comparison with ANM is ignored since it is still unknown how to extend ANM to recover
damped signals. The length of the test signal $2N-1$ is 127.}
\label{fig2}
\end{figure*}

We conduct a series of numerical experiments to examine the phase transition for exact recovery and estimate signal parameters. Superpositions of undamped and damped complex sinusoids are used as test signals. We follow the setup in \cite{Cai-RobustLR-2016, Cai-FIHT-2016} to generate the test signals. Specifically, the true signal $\bm{y} = {[{y_1}, \ldots ,{y_{127}}]^T}$ where ${y_k} = \sum\nolimits_{r = 1}^R {{c_r}{e^{i2\pi {f_r}k}}} $ and ${y_k} = \sum\nolimits_{r = 1}^R {{c_r}{e^{(i2\pi {f_r} - {\tau _r})k}}} $, $k = 1, \ldots ,127$, for undamped and damped complex sinusoids, respectively. Each frequency ${f_r}$ is drawn from the interval $[0,1)$ uniformly at random, and each complex amplitude ${c_r}$ is complex coefficients that satisfies the model ${c_r} = (1 + {10^{0.5{m_r}}}){e^{i2\pi {\theta _r}}}$ with ${m_r}$ and ${\theta _r}$ being uniformly randomly sampled from the interval $[0, 1]$. The damping parameters ${\tau _r}$ follow the model ${\tau _r} = 1/(10 + 30{n_r})$, where ${n_r}$ are uniformly randomly drawn from the interval $[0, 1]$. The signal is normalized by dividing the maximum magnitude of its entries. Then, $M$ entries of the test signals are sampled uniformly at random. For each ($R$,$M$) pair, 100 Monte Carlo trials are conducted. Each trial is declared successful if ${\left\| \bm{x} - \bm{y} \right\|}/\left\| \bm{y} \right\|\le {10^{ - 3}}$, where $\bm{x}$ and $\bm{y}$ are the true and reconstructed signals, respectively.

We plot in Fig. \ref{fig1} and Fig. \ref{fig2} the rate of successful reconstruction of undamped and damped complex sinusoids, respectively. The black and white region indicate a 0\% and 100\% of successful reconstruction, respectively, and a gray region between 0\% and 100\%. The top four plots in Fig. \ref{fig1} present the recovery phase transitions where no separation of the frequencies is imposed, while the bottom four plots present the recovery phase transitions where the wrap-around distances between the randomly drawn frequencies are greater than 1.5/$(2N-1)$.

\begin{figure*}[htb]
\centering
\includegraphics[width=6.5in]{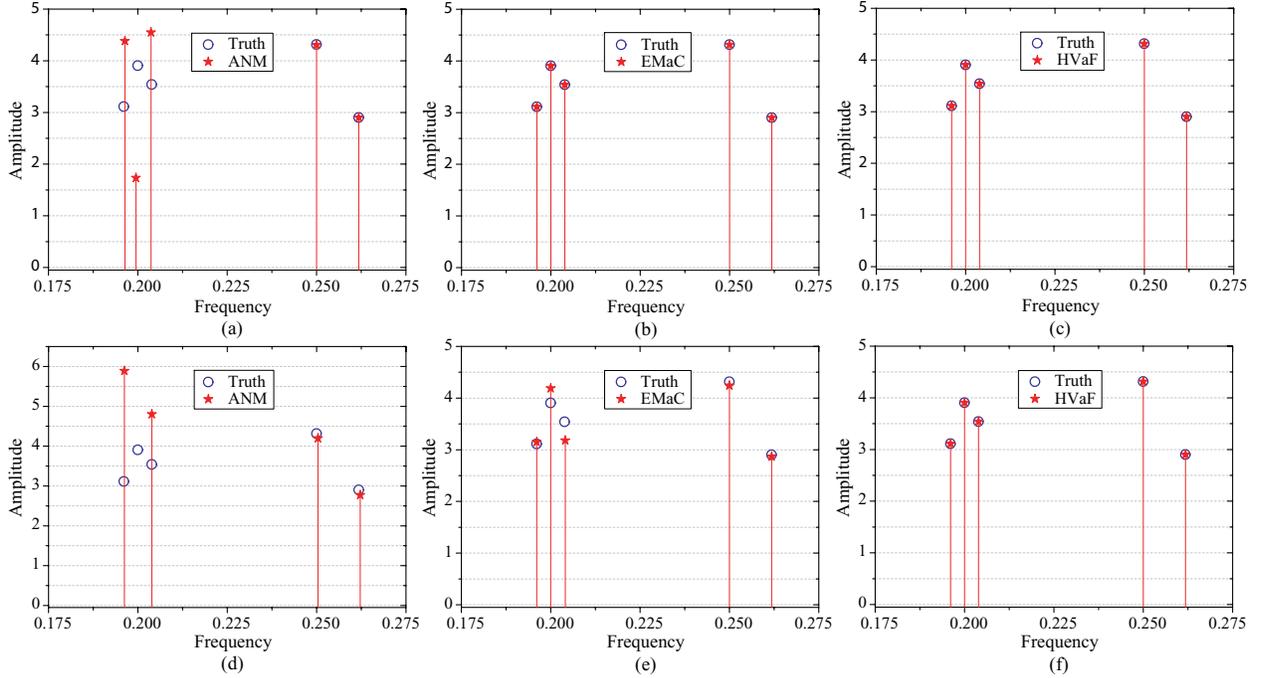}
\caption{Parameter estimation of undamped signal recovery. (a), (b) and (c) are the estimation of recovery from 50 measurements, and (d), (e) and (f) are the estimation of recovery from 25 measurements. The synthetic data is undamped complex exponential signal with 127 entries and 5 peaks. The frequencies of peaks are $0.2-0.5/127$, 0.2, $0.2+0.5/127$, 0.25, $0.25+1.5/127$, implying that the frequency separation of the three peaks in the left side and the two peaks in the right side are $0.5/(2N-1)$ and $1.5/(2N-1)$, respectively. In addition, one peak is not drawn in (d) since its estimated frequency retrieved from ANM reconstruction is beyond the scale of horizontal axis. Besides, estimation results for FIHT are omitted since the algorithm is not convergent in this specific experiment. Here $\rho$ in HVaF is 1.01.}
\label{fig3}
\end{figure*}

Fig. \ref{fig1} shows the empirical phase transitions of undamped complex sinusoid recovery for the four compared algorithms. The phase transition boundary of HVaF is significantly higher than ANM, EMaC and FITH when no separation is imposed on frequencies, implying that, for a fixed number of exponentials, HVaF requires a smaller number of measurements for successful reconstruction. While the frequencies of test signals are separated, the phase transition boundaries of ANM and HVaF are similar, slightly higher than FIHT and EMaC. Moreover, the performance of ANM and FIHT degrades severely when the separation condition is not met, while EMaC and HVaF can still achieve good performance. Therefore, HVaF and EMaC are less sensitive to the separation requirement. It is worth mentioning that the required separation condition cannot guarantee to be satisfied in practice when the number of components $R$ is relatively high.

Fig. \ref{fig2} illustrates the empirical phase transitions of damped complex sinusoid recovery for EMaC, FIHT and HVaF. This type of signals arise in NMR spectroscopy \cite{Qu-Accelerate-2015}. The comparison with ANM is ignored since it is still unknown how to extend ANM to recover damped signals. Fig. \ref{fig2} indicates that the phase transition of HVaF is much higher than EMaC and FIHT.

We further evaluate the reconstruction performance in terms of parameter estimation. In particular, we conduct the parameter estimation on the signal with small frequency separation, considering it is still challenging in this case to retrieve true parameters through reconstructions. A synthetic experiment is conducted on the signal with $5$ peaks, as shown in Fig. \ref{fig3} and ESPRIT \cite{roy1989esprit,stoica2005spectral} is used to estimate frequencies and amplitudes of the reconstructed signals. The simulated signal includes two types of frequency separations, $0.5/(2N-1)$ and $1.5/(2N-1)$. The top and bottom three plots in Fig. \ref{fig3} present the estimation of signals recovered from 50 and 25 samples, respectively.

Figs. \ref{fig3}(a) and (d) show that the amplitudes of the three peaks on the left side, where the frequency separation is $0.5/(2N-1)$, are estimated much worse than these of the two peaks on the right side, where the frequency separation is $1.5/(2N-1)$. It is observed that the true parameters with small frequency separation are not retrieved from ANM reconstruction. Figs. \ref{fig3}(b) and (e) show that the parameters with small frequency separation can be estimated accurately through EMaC reconstruction from 50 measurements, while estimation error increases when the number of measurements decreases to 25, implying that a relatively high sampling rate is necessary for EMaC to retrieve true parameters. It is observed from Fig. \ref{fig3}(c) and (f) that HVaF presents accurate parameter estimations in the case of small frequency separation and low sampling rate.

\begin{figure*}
  \begin{minipage}{0.49\textwidth}
  \centering
\includegraphics[width=2.8in]{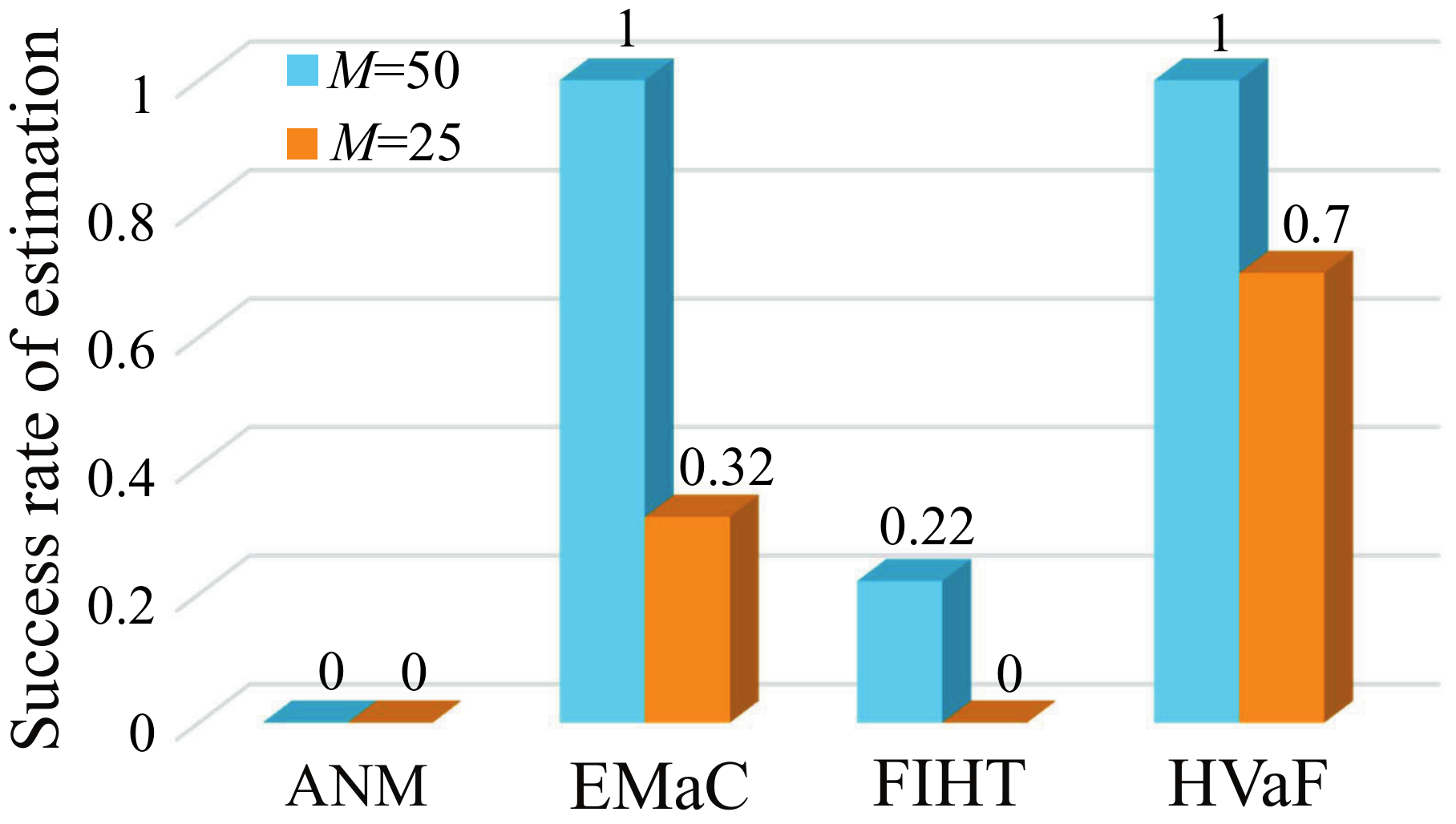}
\caption{Success rate of parameter estimation of undamped signal recovery. The success rate is calculated over 100 Monte Carlo trials.}
\label{fig4}
\end{minipage}
\hfill
\begin{minipage}{0.49\textwidth}
\centering
\includegraphics[width=2.5in]{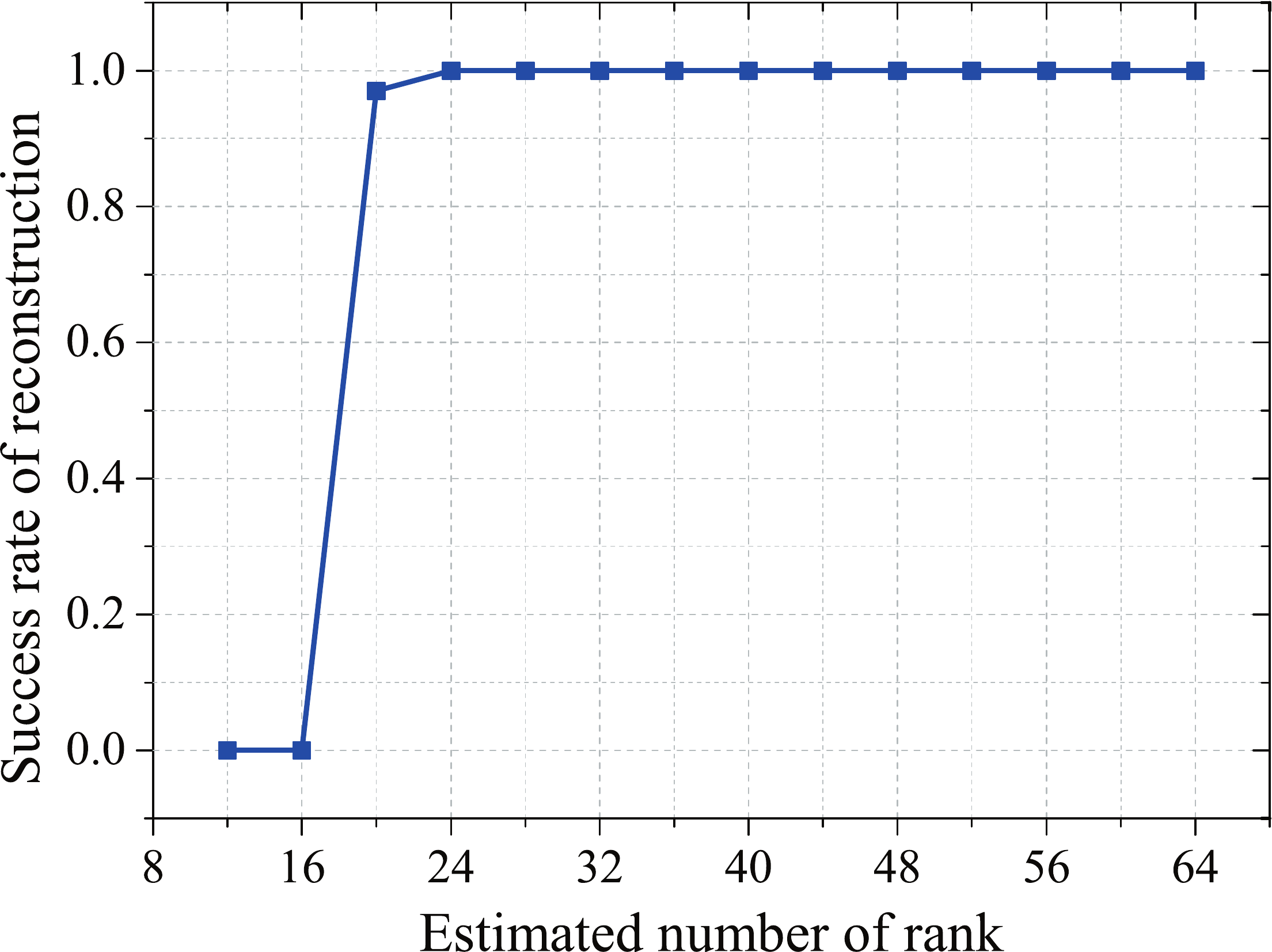}
\caption{The success rate of reconstructions versus estimated rank of HVaF. The exact number of exponentials $R$ is $20$ and the number of measurements $M$ is $80$.}
\label{fig5}
\end{minipage}
\end{figure*}

We further conduct 100 Monte Carlo trials of parameter estimation for the synthetic signal in Fig. \ref{fig3} and each trial is declared successful if $\sqrt {\sum\nolimits_{r = 1}^R {\left( {\hat f}_r - {f_r} \right)}^2 } /\sqrt {\sum\nolimits_{r = 1}^R {f_r^2} }  \le {10^{ - 3}}$ and $\sqrt {\sum\nolimits_{r = 1}^R {{\left( {\left| {\hat c}_r \right| - \left| c_r \right|} \right)}^2} } /\sqrt {\sum\nolimits_{r = 1}^R {\left| {{c_r}} \right|^2} }  \le {10^{ - 3}}$ are simultaneously satisfied, where $\hat f$ and ${\hat c}_r$ are the estimated frequency and amplitude of the true $f_r$ and $c_r$. Fig. \ref{fig4} shows that HVaF achieves higher success rate of parameter estimation than ANM and FIHT when frequency separation is small. Compared with EMaC, HVaF requires less measurements to obtain successful parameter estimation.

Fig. \ref{fig5} evaluates the effect of preset number of exponentials $\hat R$ on reconstruction. It is observed that HVaF can always achieve high success rate as $\hat R$ increases from 20 to 64, while the exact number of exponentials is 20. This observation indicates that HVaF may have great potential to reconstruct a signal even in the case that a much larger $\hat{R}$ than $R$ is given. Thus, flexible setting of $\hat{R}$ is possible for the proposed HVaF.

Table \ref{tab_time} shows the average computational time of each method. We can see the large estimated rank $\hat R$ in HVaF will result in the long computational time. Therefore, $\hat R$ should not be overrated too much considering the computational cost, though the reconstruction results will not be degraded.

\begin{table*}[htbp]
\centering
\caption{Comparisons of empirical computational time (sec)}\label{tab_time}
\begin{tabular}{|c|c|c|c|c|c|}
\hline
ANM&EMaC&FIHT&HVaF ($\hat{R}=10$)&HVaF ($\hat{R}=30$)&HVaF ($\hat{R}=50$)\\
\hline
5.05&54.87&0.16&21.16&168.94&292.77\\
\hline
\end{tabular}
\begin{tablenotes}
  \footnotesize
  \item Note: The synthetic undamped signal consists of 10 peaks. The length of the signal is 127 and the number of measurements is 64. The computational time is computed by averaging $100$ Monte Carlo trials. The numerical experiments are conducted on a Dell PC running Windows 10 operating system with Intel Core i5 4570 CPU and 8-GB RAM.
\end{tablenotes}
\end{table*}

\subsection{Real NMR spectroscopy data}\label{sec:NMR}
NMR spectroscopy plays an important role in studying structure, dynamics and interactions of biopolymers in chemistry and biology. The non-uniform sampling is popular to reduce the number of measurements \cite{Hyberts2010, Orekhov2011, Qu-Accelerate-2015, Qu-sparse-2010, Kazimierczuk2011, holland2011fast, Qu2011} due to the long duration of spectroscopy experiment.

The time domain signal of NMR is usually modelled as a superposition of damped complex sinusoids \cite{Qu-Accelerate-2015, hoch1996nmr}. The efficiency of LRHM has been verified in NMR spectroscopy \cite{Qu-Accelerate-2015}, showing advantages in recovering broad peaks over ${l_1}$ norm minimization on the spectrum \cite{Qu-sparse-2010, Kazimierczuk2011, holland2011fast, Qu2011}. However, LRHM needs a relatively high sampling rate, for example 35\% for the following NMR spectrum \cite{Qu-Accelerate-2015}, to obtain a reliable reconstruction. In this section, we will compare HVaF with LRHM in recovering realistic biological NMR spectroscopy data with a lower sampling rate. The comparisons with ANM and FIHT are ignored, since ANM is not available in recovering damped signals and the number of exponentials in realistic  NMR spectroscopy cannot be exactly estimated for FIHT.

Here we apply HVaF to recover a $^1\mathrm{H}{ - ^{15}}\mathrm{N}$ spectrum from Poisson-gap \cite{Hyberts2010} non-uniformly sampled time-domain data. The data compose a matrix $\bm X$ with the size $255 \times 116$, where columns and rows indicate $^{15}\mathrm{N}$ and $^{1}\mathrm{H}$ dimensions, respectively. According to the principle of NMR experiments, each column of the spectrum is non-uniformly sampled and can be reconstructed individually. The spectrum is normalized by dividing the maximum magnitude of its entries as pre-processing.

\begin{figure}[htb]
\centering
\includegraphics[width=3.4in]{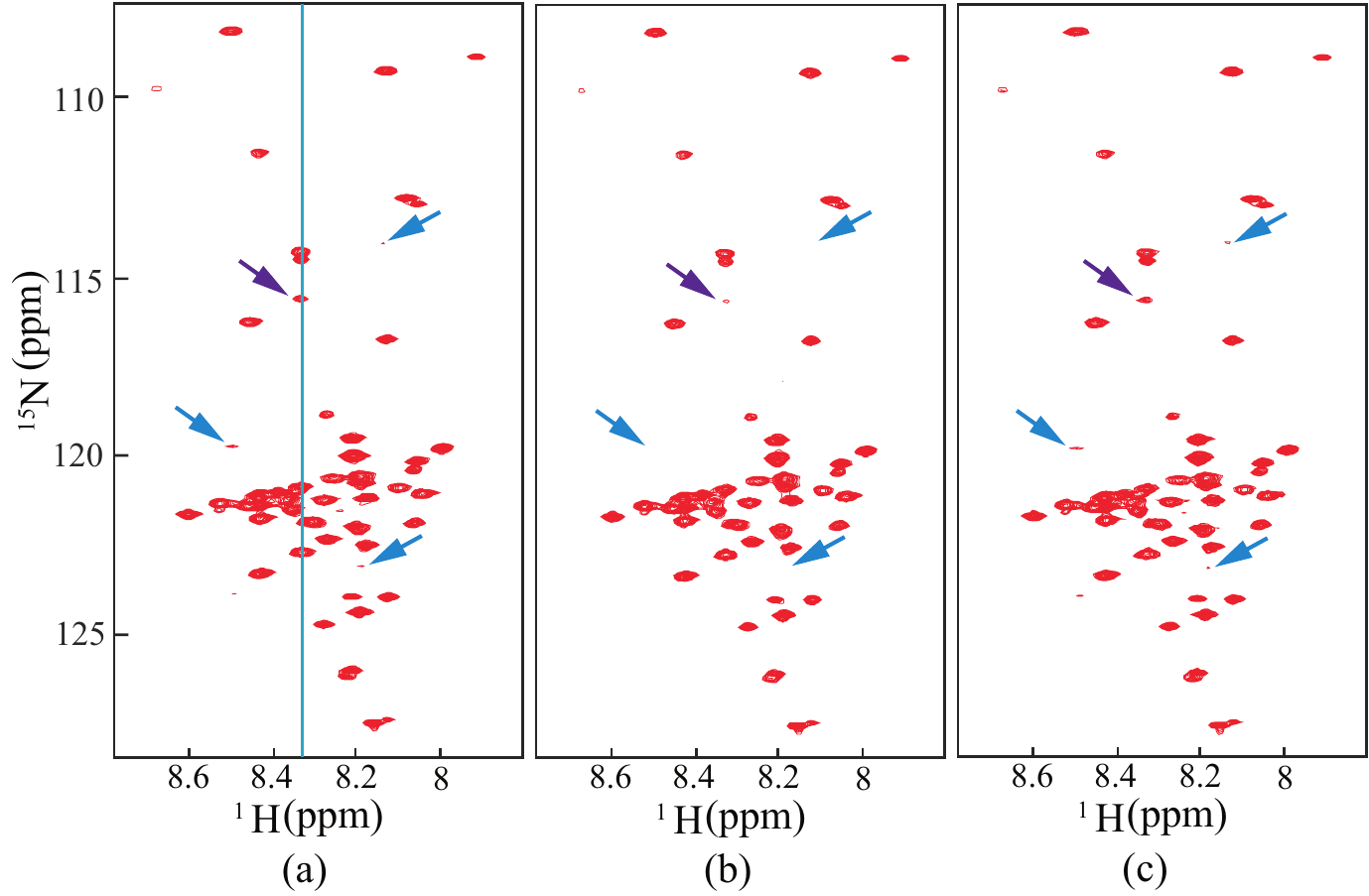}
\caption{The NMR spectra recovery under 22\% non-uniform sampling. (a) Fully sampled spectrum; (b) and (c) are the reconstructed spectra by LRHM and HVaF, respectively. The ppm denotes parts per million by frequency, the unit of chemical
shift.}
\label{fig6}
\end{figure}

\begin{figure}[htb]
\centering
\includegraphics[width=3.4in]{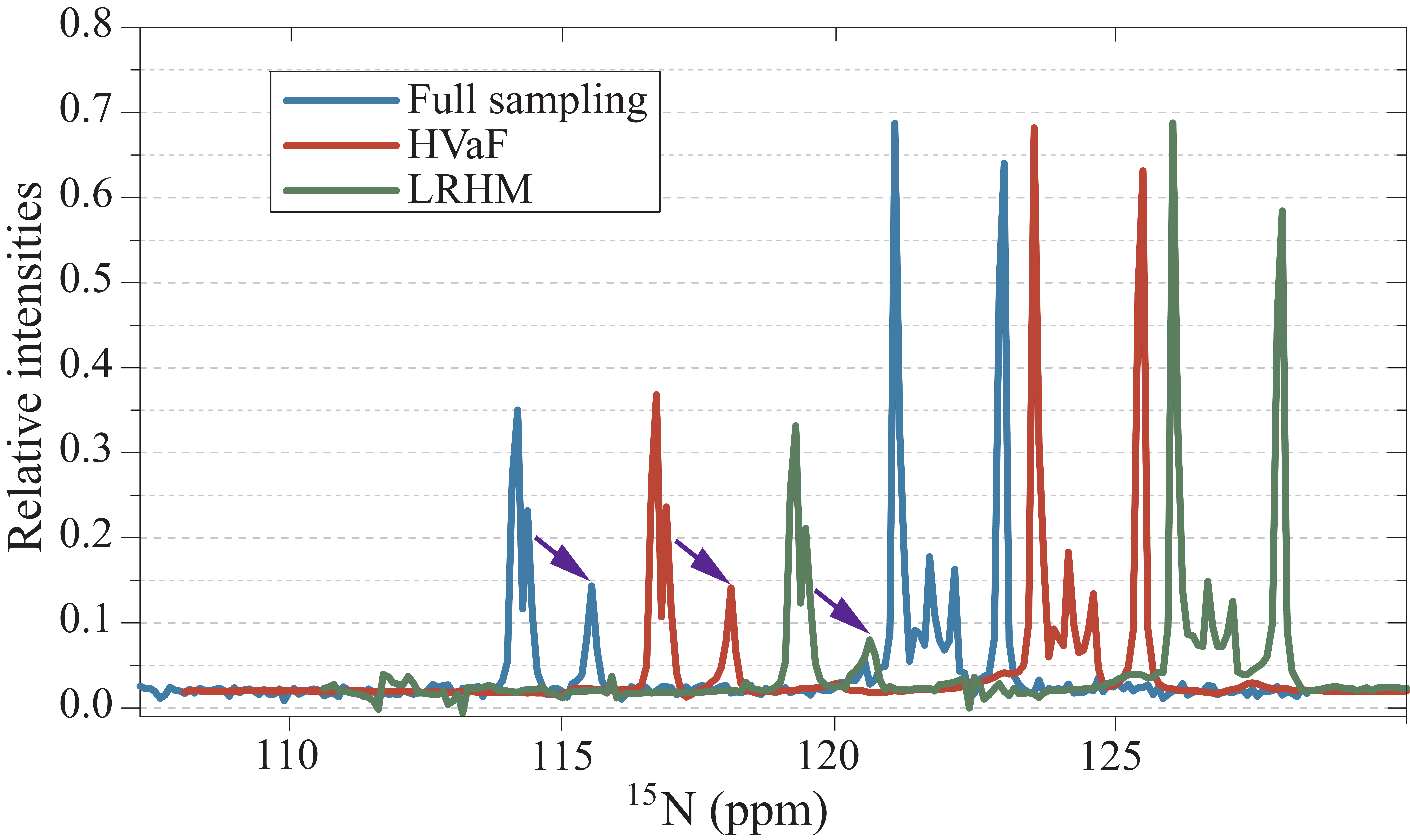}
\caption{One column of the reconstructed spectra by LRHM and HVaF. The column spectrum is along the dimension of $^{15}\mathrm{N}$ and are located at 8.35 ppm of the dimension of $^{1}\mathrm{H}$. Note: The 1D spectrum are shifted for better visualization.}
\label{fig7}
\end{figure}

\begin{figure}[htb]
\centering
\includegraphics[width=3.5in]{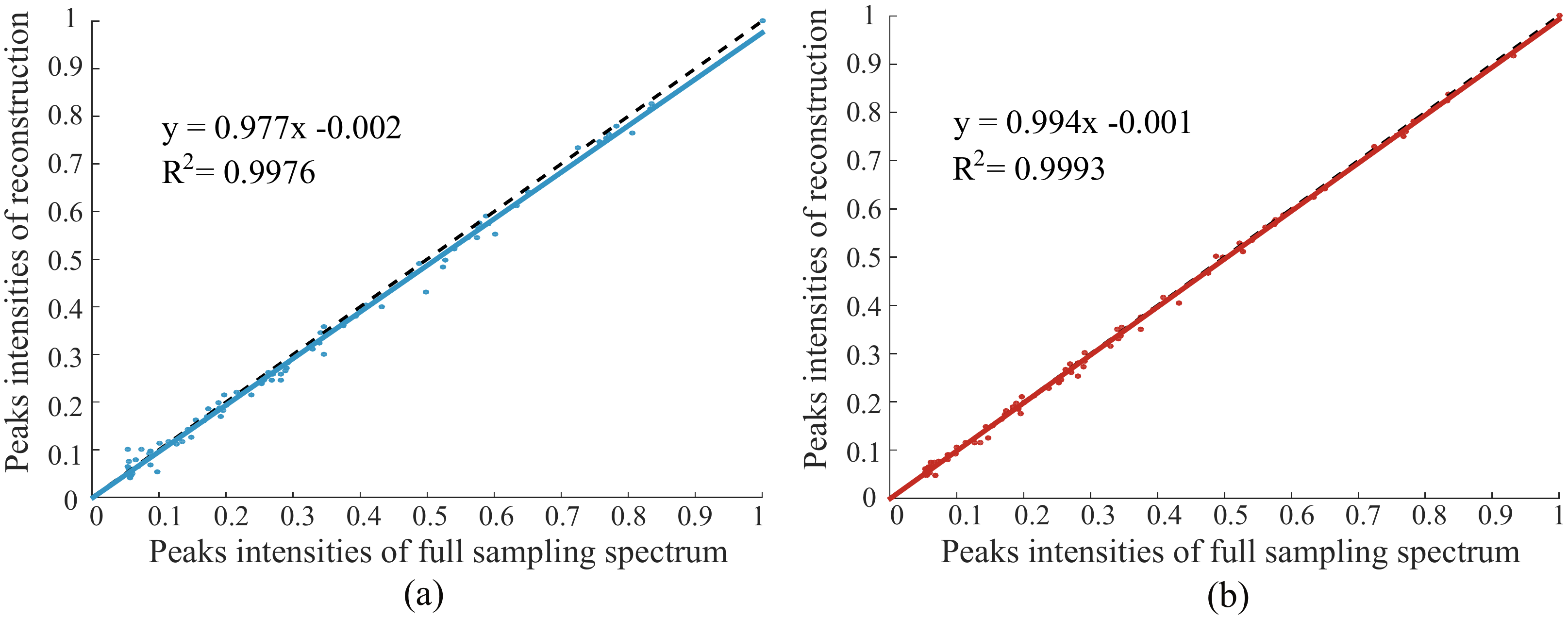}
\caption{Peak intensities correlation between the full sampled spectrum and the reconstructed spectrum. (a) and (b) are correlation evaluation for LRHM and HVaF, respectively. Note that the notation $\mathrm{R}^2$ denotes Pearsons linear correlation coefficient of fitted curve. The closer the value of $\mathrm{R}^2$ gets to 1, the stronger the correlation between the full sampled spectrum and the reconstructed spectrum is. Here, 83 peaks are extracted and their intensities, obtained by finding the local maximum within a spectrum region (three by three points), are adopted in correlation analysis. Spectrum intensities that are smaller than the noise level, 0.05 in this case, are treated as noise and will not be plotted or used for correlation analysis.}
\label{fig8}
\end{figure}

Fig. \ref{fig6} shows the reconstructed spectra using a 22\% sampling rate, indicating that HVaF leads to a more faithful recovery of the full sampled spectra than LRHM. As marked by the arrows in Fig. \ref{fig6}, some peaks are underestimated severely in LRHM reconstruction but not in HVaF. Here we compute the Relative Least Normalized Error (RLNE) by
\begin{equation}
\Vert \bm{X}-\bm{Y}\Vert_F / \Vert \bm{Y}\Vert_F,
\end{equation}
where $\bm{X}$ is the reconstructed 2D spectrum and $\bm{Y}$ is the fully sampled 2D spectrum with noise. The reconstruction errors  by HVaF and LRHM are 0.1036 and 0.1127, respectively. But note that the reconstruction errors here are not very conclusive, since the fully sampled spectrum is noisy. Therefore we further present qualitative results and peak corrections. Fig. \ref{fig7} illustrates one column of the reconstructed spectra. It is observed that HVaF achieves a reliable reconstruction while LRHM weakens the marked peak and introduces some artifact peaks. Fig. \ref{fig8}  further presents that the HVaF achieve higher accuracy of peak intensities. This observation is consistent with Fig. \ref{fig2} showing that a higher success rate is obtained with the proposed approach under the same number of measurements. The advantage of HVaF over LRHM implies a more significant reduction in measurement time and thus HVaF will be valuable for the biological NMR applications.

\section{Discussions}\label{sec:V}

\subsection{Parameter setting in the model and algorithm}

\begin{figure*}[htp]
  \centering
 \includegraphics[width=0.98\textwidth]{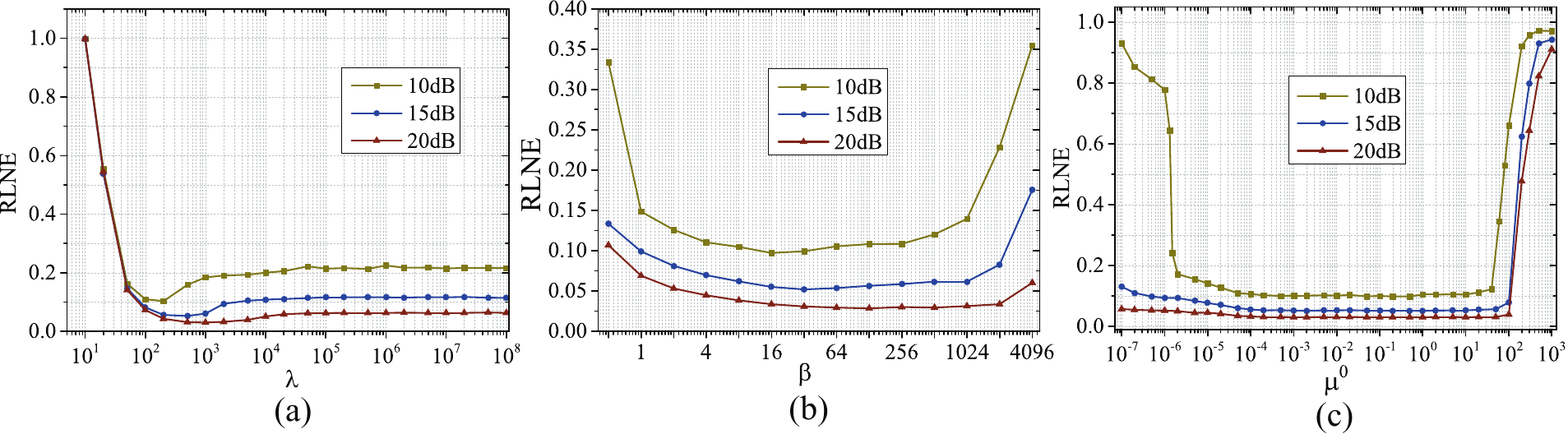}
 \caption{RLNE versus parameters (a) $\lambda$, (b) $\beta$ and $\mu^0$ under different noise levels. In (a), $\beta$ and $\mu^0$ are 32 and 0.05, respectively; In (b), the $\lambda$ is set as 200, 500 and 1000 under noise levels 10dB, 15dB and 20dB, respectively. $\mu^0$ is 0.05; In (c), the setting of $\lambda$ is the same with that in (b) and $\beta$ is 32.}\label{fig_para}
\end{figure*}

The proposed method includes the parameter $\lambda$ in the mathematical model as well as $\beta$ and $\mu^0$ in the numerical algorithm. Fig. \ref{fig_para} shows the impact of $\lambda$, $\beta$ and $\mu^0$ on the reconstruction error, respectively. The synthetic signal in these experiments consists of $5$ complex exponentials and the number of observed entries is $64$. The measurements are corrupted by the noise
\begin{equation}
\bm e =\sigma \cdot \|\mathcal{P}_\Omega(\bm y)\| \cdot \frac{\bm w}{\|\bm w\|},
\end{equation}
where $\| \cdot \|$ is $\mathcal{\ell}_2$-norm, $\bm y$ is the true signal, the entries of $\bm w$ are i.i.d. standard Gaussian random variables and $\sigma$ is referred to as the noise level. The $\sigma$ is set as $10^{-0.5}$, $10^{-0.75}$ and $10^{-1}$ in experiments, and the corresponding signal-to-noise ratios (SNR), is $10$dB, $15$dB and $20$dB, where SNR is computed by
\begin{equation}
{\rm SNR} =-10\log_{10} \frac{\|\bm e\|^2}{\| \mathcal{P}_\Omega (\bm y)\|^2}.
\end{equation}

Fig. \ref{fig_para}(a) indicates that the available range, leading to the reconstruction error ${\rm RLNE} \leq 0.1$, turns narrowed when the noise level increases. The optimal $\lambda$, which produces the lowest reconstruction error, generally decreases as the noise level increases. For example, as shown in Fig. \ref{fig_para}(a), the optimal $\lambda$  is 200, 500 and 1000 for noise levels of $10$dB, $15$dB and $20$dB, respectively. This means that a smaller $\lambda$ should be set for a higher noise level. From Figs. \ref{fig_para}(b) and \ref{fig_para}(c), we can see the reconstruction error is steady over a large range of choice of $\beta$ and $\mu^0$ under different noise levels.

\subsection{Parameter estimation of damped signal recovery}

Experiments on parameter estimation of damped signal recovery are conducted here.

It is observed that all of the compared methods, EMaC, FIHT and HVaF, achieve accurate parameter estimation if sufficient number of samples, 60 samples in the simulation, are available. However, the performance of EMaC degrades severely when the number of samples is reduced to 30 as shown in Table \ref{Estimation}, while HVaF still obtains good performance. This observation implies that HVaF requires fewer samples to achieve accurate estimation than EMaC.

\begin{table*}[ht]
\centering
\caption{Parameter estimation comparisons of damped signal recovery.}
\label{Estimation}
\begin{tabular}{|c|c|c|c|c|}
\hline
\multirow{2}{*}{ Peak index} & \multirow{2}{*}{\tabincell{c}{True parameters \\ $(|c_r|, f_r, \tau_r)$}} & \multicolumn{2}{c|}{Parameter estimation errors }        \\ \cline{3-4}
                  &                                                                 & EMaC                     & HVaF ($10^{-6}, 10^{-8}, 10^{-4} $)\\ \hline
1                 & 0.5145, 0.1532, 26.47                                    &   0.2097, 0.1603, 26.25     &   2.8, \ 1.4, \ 1.1  \\ \hline                                                                 
2                 & 0.6623, 0.3135, 35.63                                    &   0.3615, 0.1594, 0.437      &   2.3, \ 2.0, \ 0.4  \\ \hline                                                                     
3                 & 0.7253, 0.4716, 48.78                                     &   0.2157, 0.0001, 14.34   &   2.7, \ 0.2, \ 1.3  \\ \hline                                                                  
4                 & 0.7825, 0.6124, 61.51                                    &   0.2060, 0.0003, 19.63    &   0.7, \ 0.5, \ 0.4 \\ \hline                                                                    
5                 & 0.9872, 0.7831, 81.50                                    &   0.1531, 0.0003, 14.75   &   1.5, \ 0.3, \ 1.2   \\ \hline                                                                  
\end{tabular}
\begin{tablenotes}
\footnotesize
  \item Note:The synthetic signal with 5 peaks is generated by ${y_k} = \sum\nolimits_{r = 1}^5 {{c_r}{e^{(i2\pi {f_r} - {\tau _r})k}}} $, $k = 1, \ldots ,127$, where $c_r$, $f_r$ and $\tau_r$ denote the complex amplitude, frequency and damping factor of the $r^{\rm{th}}$ peak, respectively.The number of samples is $30$. ESPRIT \cite{roy1989esprit,stoica2005spectral} is used to estimate the parameters of the reconstructed signals. The results of FIHT are omitted since FIHT is not convergent in this case.
\end{tablenotes}
\end{table*}

\subsection{Frequency identifiability}

The results shown in  Fig. \ref{fig1} imply that the proposed method achieved a better signal reconstruction performance when frequencies were closer. A possible reason is that HVaF favors deciding that there is a single one frequency component in the case that two frequencies are extremely close with each other.

This guess is further confirmed by the results shown in Table \ref{tab_EST}, indicating that the method is not able to distinguish frequency components that are close together in the missing data recovery. Table \ref{tab_EST} shows an experiment on two-component signal reconstruction. Here we consider two cases of frequency separations between the two frequency components, $0.01/(2N-1)$ and $1.5/(2N-1)$. It is observed that, to obtain low reconstruction error(RLNE $\leq 10^{-3}$), the signal with frequency separation $0.01/(2N-1)$ requires less measurements than the signal with separation $1.5/(2N-1)$. Spectral parameter estimation on the reconstructed signal shows that, the original two close peaks are synthesized together since the magnitude of one spectral peak has been reduced from 0.66 to $1.4 \times 10^{-4}$. This observation implies that the rank of the Hankel matrix in the iterative reconstruction may be reduced to 1. Thus, a small number of measurements is possible in the reconstruction.

However, our method is intended to reconstruct the missing data rather than estimate spectral parameters, and thus aims to obtain a low signal reconstruction error. In summary, the proposed HVaF has shown much better missing data reconstruction performance than the compared methods. The HVaF still has limitation on preserving very close frequencies in the case of missing data, which is always very challenging and will be a valuable future work.

\begin{table*}[htp]
\centering
\caption{Parameter estimation under different frequency separations.}
\label{tab_EST}
\begin{tabular}{|c|c|c|c|c|c|c|}
\hline
\multirow{2}{*}{Peak index} & \multicolumn{2}{c|}{True parameters $f_r$, $|c_r|$}
                  &    \multicolumn{2}{c|}{ Estimated $\hat{f}_r$, $|\hat{c}_r|$ ($M=8$) }      &  \multicolumn{2}{c|}{Estimated $\hat{f}_r$, $|\hat{c}_r|$ ($M=12$)}        \\ \cline{2-7}
                  & 0.01/(2$N$-1)&1.5/(2$N$-1) &0.01/(2$N$-1)&1.5/(2$N$-1)&0.01/(2$N$-1)&1.5/(2$N$-1) \\\hline
 1 & 0.3000, 0.5100&0.3000, 0.5100 & 0.3000, 1.1681 & 0.2992, 0.4346 & 0.3000, 1.1672 & 0.3000, 0.5100 \\ \hline
 2 & 0.3001, 0.6600 & 0.3118, 0.6600 & 0.2981, $1.4\times 10^{-4}$ & 0.3328, 0.3550 & 0.3004, $9.8\times 10^{-5}$ & 0.3118, 0.6600 \\ \hline
\multicolumn{3}{|c|}{Reconstruction error (RLNE)} &$3.7\times10^{-5}$ &$0.2163$ &$1.0\times10^{-5}$ &$5.6\times10^{-6}$  \\ \hline
\end{tabular}
\begin{tablenotes}
  \footnotesize
  \item Note:The synthetic signal with 2 peaks is generated by ${y_k} = \sum\nolimits_{r = 1}^2 {{c_r}{e^{2\pi i {f_r}k}}} $, $k = 1, \ldots ,127$, where $c_r$ and $f_r$ denote the complex amplitude, and frequency of the $r^{\rm{th}}$ peak, respectively. ESPRIT \cite{roy1989esprit,stoica2005spectral} is used to estimate the parameters of the reconstructed signals.
\end{tablenotes}
\end{table*}

\section{Conclusions}\label{sec:VI}

A new approach, HVaF, is proposed to reconstruct the exponential signal by exploiting the Vandermonde factorization of the Hankel matrix formed by the signal. To implement HVaF, a numerical algorithm was developed and the sequence convergence were analysed. Experiments on synthetic data demonstrated that HVaF achieves higher empirical phase transition than nuclear-norm based minimization and fast iterative hard thresholding algorithm. Another advantage over state-of-the-art atomic norm minimization and fast iterative hard thresholding algorithm was empirically observed that HVaF can reconstruct signals with small frequency separation. The proposed method was further verified on reconstruction of fast sampled NMR spectroscopy, implying that HVaF may serve as an effective method for fast sampling of NMR spectroscopy in chemistry and biology.

For future work, it is of great interest to develop more efficient numerical methods to solve HVaF when the datasets are huge. In addition, we can also introduce rank minimization \cite{le2018light}, truncated nuclear norm \cite{oh2016partial}, or weighted nuclear norm \cite {guo2018improved} which may achieve extra improvements in reconstruction, e.g. better recovery of low intensity spectral peaks, since the low rank property can be better approximated than the commonly used nuclear norm. The code will be available at http://csrc.xmu.edu.cn.

\section{Acknowledgments}
The authors thank Yuejie Chi for sharing EMaC codes for comparisons and thank Vladislav and Maxim for sharing the NMR data used in paper \cite{Qu-Accelerate-2015}. J. Ying would like to thank Benjam\'in B\' ejar Haro and Qiuwei Li for helpful discussions related to this paper. The authors also appreciate reviewers for their constructive comments.

\section{Appendix}
\subsection{The closed-form solutions of $\bm{U}$ and $\bm{V}$}

The closed-form solution of $\bm{U}$ can be obtained by solving \eqref{eq:SolU}.
According to definitions of the operators $\mathcal{R}$ and $\mathcal{R}^*$, we obtain $\mathcal{R}^*\mathcal{R}\bm{x} = \bm{w} \odot \bm{x}$, where $\bm{w}$ is a vector and its $k$-th element ${w_k}$ is the number of elements in $k$-th anti-diagonal of the Hankel matrix $\mathcal{R}\bm{x}$. Here $ \odot $ denotes Hadamard product. In addition, according to definitions of $\mathcal{Q}$ and $\mathcal{Q}_r^*$, we have $\left[ \mathcal{Q}_r^*\mathcal{Q}_r\bm{X} \right]_{(:,k)} = \left\{ {\begin{array}{*{20}{c}}
\bm{X}_{(:,r)}\\
0
\end{array}{\kern 10pt} \begin{array}{*{20}{c}}
{k = r,}\\
{k \ne r.}
\end{array}} \right.$ Hence, for $\bm{X} \in \mathbb{C}^{N \times N}$, by combination we get
\begin{equation*}
\sum\limits_{r = 1}^{\hat R} \mathcal{Q}_r^*\mathcal{R}^*\mathcal{R}\mathcal{Q}_r\bm{X} = \bm{T} \odot \bm{X},
\end{equation*}
where $\bm{T}$ is a constant matrix and each column of $\bm{T}$ is $\bm{w}$. Therefore, we rewrite \eqref{eq:SolU} as
\begin{equation*}
\mu ^k\bm{T} \odot \bm{U} + \beta\bm{U}(\bm{V}^k)^T\mathrm{conj}(\bm{V}^k) = \bm{Y}.
\end{equation*}
where the right term of \eqref{eq:SolU} is denoted by $\bm{Y}$. Obviously, we can obtain the closed-form solution of each row of $\bm{U}$ by
\begin{equation*}
\mu ^k\bm{U}_{(r,:)}\bm{T}_r + \beta\bm{U}_{(r,:)}(\bm{V}^k)^T\mathrm{conj}(\bm{V}^k) = \bm{Y}_{(r,:)},
\end{equation*}
and its solution is
\begin{equation*}
\bm{U}_{(r,:)} = \bm{Y}_{(r,:)}(\mu^k\bm{T}_r + \beta(\bm{V}^k)^T\mathrm{conj}(\bm{V}^k))^{ - 1}.
\end{equation*}
where $\bm{T}_r$ is a diagonal matrix and its main diagonal is $\bm{T}_{(r,:)}$.

Therefore we can update $\bm{U}$ by updating each row and it is similar to update $\bm{V}$.

\ifCLASSOPTIONcaptionsoff
  \newpage
\fi

\bibliographystyle{IEEEtran}
\bibliography{MyBibliography}

\end{document}